\PassOptionsToPackage{bookmarksnumbered,unicode,colorlinks=true,linktocpage=true}{hyperref}
\documentclass[a4paper,fleqn]{cas-sc}

\usepackage[numbers]{natbib}

\def\tsc#1{\csdef{#1}{\textsc{\lowercase{#1}}\xspace}}
\tsc{WGM}
\tsc{QE}
\tsc{EP}
\tsc{PMS}
\tsc{BEC}
\tsc{DE}

\usepackage{url}
\usepackage[nameinlink,capitalize]{cleveref}
\usepackage{hyperref}
\usepackage{subfig}
\graphicspath{{Figures_final//}}
\DeclareGraphicsExtensions{.eps,.pdf,.png,.jpg,.tif,.tiff,.ps}

\usepackage{graphics} %
\usepackage{epsfig} %
\usepackage{float} %
\usepackage{graphicx}

\usepackage[justification=centering]{caption}

\usepackage{stix}

\usepackage{enumitem}

\usepackage{xspace}
\newcommand{\ti}{\textsc{Ti}\xspace}

\newcommand{\eat}[1]{}

\newcommand{\revA}[1]{{\color{black}{#1}}}

\newcommand{\verify}[1]{{\color{black}{#1}}}

\begin{document}
\let\WriteBookmarks\relax
\def\floatpagepagefraction{1}
\def\textpagefraction{.001}
\shorttitle{Traffic prediction using Deep Learning}
\shortauthors{Mihaita et~al.}

\title [mode = title]{Traffic congestion anomaly detection and prediction using deep learning}                      

\author[1]{Adriana-Simona Mihaita}
\cormark[1]
\fnmark[1]
\ead{adriana-simona.mihaita@uts.edu.au}
\ead[url]{www.fmlab.org}
\ead{Declarations of interest: none.}
\credit{Conceptualization of this study, Methodology}
\address[1]{University of Technology Sydney, 61 Broadway Str, Sydney, Australia}

\author[2]{Haowen Li}%
\address[2]{Australian National University, Canberra, Australia}

\author[1]{Marian-Andrei Rizoiu}
\credit{Methodology, Deep Learning framework}

\cortext[cor1]{Corresponding author}

\begin{abstract}
Congestion prediction represents a major priority for traffic management centres \revA{around the world} to ensure timely incident response handling. 
\revA{The increasing amounts of generated traffic data have been used to train machine learning predictors for traffic, however this is a challenging task} due to inter-dependencies of traffic flow both in time and space. 
Recently, deep learning techniques have shown significant \revA{prediction} improvements over traditional models, however open questions remain around their applicability, accuracy and parameter tuning. 
\verify{This paper bring two contributions in terms of: 1) applying an outlier detection an anomaly adjustment method based on incoming and historical data streams, and 2) proposing an advanced deep learning framework for simultaneously predicting the traffic flow, speed and occupancy on a large number of monitoring stations along a highly circulated motorway in Sydney, Australia, including exit and entry loop count stations, and over varying training and prediction time horizons.
The spatial and temporal features extracted from the 36.34 million data points are used in various deep learning architectures that exploit their spatial structure (convolutional neuronal networks), their temporal dynamics (recurrent neuronal networks), or both through a hybrid spatio-temporal modelling (CNN-LSTM).
We show that our deep learning models consistently outperform traditional methods, and we conduct a comparative analysis of the optimal time horizon of historical data required to predict traffic flow at different time points in the future. Lastly, we prove that the anomaly adjustment method bring significant improvements of using deep learning in both time and space. 
}
\end{abstract}


\maketitle

\section{INTRODUCTION}\label{S1_Intro}

Traffic congestion represents one of the sensitive points of many traffic management centres around the world, which need to ensure that travel times remain within regular patterns, and that incidents are cleared in due time on a daily basis. 
Predicting the dynamics of the traffic congestion within the next 30min to one hour represents a high priority for real-time traffic operations. 
The main advantage of using advanced traffic flow prediction \revA{techniques} lies in \revA{their} ability to quickly adapt to stochastic incidents, and to predict their impact starting from only incipient measurements. 
The increasing amount of traffic data generated by intelligent transport systems led to the development of multiple data-driven approaches and prediction models.
However, there are several open questions concerning traffic flow prediction:
a) how to efficiently predict road traffic congestion using extensive data-driven techniques which can adapt to real-time big-data sets?, 
b) what are the best techniques that can capture the spatial-temporal correlations arising in complex traffic networks? and 
c) why are some models efficient for short-term traffic prediction, but not for long-term prediction? 

The parametric approaches were typically based on time-series analysis, Kalman models, etc. For example, the ARIMA model has been widely applied for traffic flow prediction (\cite{VANDERVOORT1996307}, \cite{dKamarianakis2003}) due to its simplicity and good performance in forecasting linear and stationary time-series. 

Further extensions of ARIMA have been proposed to account for seasonal features (SARIMA \cite{quteprints63176}), and for additional explanatory variables (ARIMAX \cite{Williams2001}). %
The effectiveness of parametric models can be affected by the traffic stochasticity, and by the occurence of disruptive events.
As a result, non-parametric models have seen an increasing popularity, and among them we cite: k-nearest neighbours \cite{Chang2012}, support vector regressions \cite{Jeong2013}, artificial neural networks \cite{KARLAFTIS2011}, and Gaussian Processes \cite{IDe2009}.

Recently Deep Learning (DL) methods have emerged as popular non-parametric alternative approaches for short-term predictions, with various models being proposed and tested in different set-ups. 
Two major literature reviews on DL models can be found in \cite{Ali2018}, and \cite{Nguyen2018}.
These debate how different DL models can be adapted for traffic flow prediction, and why the spatial and temporal correlation in traffic congestion propagation makes the application of such models difficult. 
More recently, recurrent neural networks (RNNs) -- such as the long short-term memory model (LSTM) -- have been designed to learn from sequences of data, and to capture long-term temporal patterns.
LSTM was applied to traffic flow data~\cite{TIAN2018297}, either alone of jointly with convolutional neuronal networks (CNN) \cite{Nguyen2019,Wu2019} in an attempt to capture the spatial road network information.
Among the difficulties of deploying such models are their often complicated structure, the choice of parameters (such as the number of neurons or the non-linear functions), and the fact that neural networks have been long time regarded as ``black-boxes'' \cite{ZHANG201465}.
These difficulties are starting to ease due to the emergence of integrated modelling and fitting frameworks, such as TensorFlow~\cite{tensorflow2015} and PyTorch~\cite{paszke2017automatic}.

There are several open questions relating to the usage of DL methods for traffic flow prediction, still not addressed in the literature.
The first question relates to the scalability of such methods.
The majority of existing studies concentrate on one or several stations, or over short periods of time~\cite{Tan2018}.
It is therefore unclear how DL models behave at the level of an entire motorway, or for large
datasets with complicated road structures. 
We address this question by
constructing deep learning models capable of predicting the real-time traffic flow along an entire motorway.
Our dataset spans over an year, and it spreads across 208 stations along 48kms of road network in Sydney, Australia. 

The second open question relates to the relationship between the training and the prediction horizons.
The majority of existing work performs traffic flow prediction for future time horizons of 2-3 time intervals~\cite{POLSON20171}
however they usually fix the past horizon to 30 or 40 minutes.
This arbitrary choice can affect the prediction performances, and impact the model selection. 
In our study we propose a sensitivity analysis between various past and future horizons for each of the deep learning models under investigation, and we showcase the best set-up for each model.
We find that for LSTM there is a limit in the past horizon beyond which the accuracy stops improving, and that CNN performances are actually hurt by using too much data from the past.

The third open question is about deploying hybrid deep learning models that combine both spatial and temporal modelling.
While some research has shown that hybrid DL architectures can improve performances in specific circumstances \cite{Tan2018,Wang2016TrafficSP}, other studies still debate the necessity and overhaul of fine-tuning such models. In this paper, we implement a hybrid CNN-LSTM model and we find that it under-performs the LSTM model, with its performances fluctuating significantly with respect to the future time horizon.  

\verify{The fourth and last question is regarding the efficiency of such models under outliers and anomalies that can affect their accuracy and application. Anomaly detection methods have been widely used in various domains, but in transportation majority of work is reporting towards the usage of surveillance video cameras \citep{Bouindour2019} or combination of machine learning and video-camera streaming \citep{Djenouri2018}. Other methods have been proposed and developed separately \citep{CUADRASANCHEZ20151}, but very few in combination with deep learning algorithms or integrating the traffic behaviour revealed by the fundamental diagram of speed-flow for example. We address this question by proposing an outlier detection methods for all incoming streams in the data set (flow, speed, occupancy) and an anomaly improvement/repairing methodology which adjusts parameters not only based on historical data profiles but also on incoming data streams.

The objective of this paper is to construct an advanced deep learning framework for predicting the traffic congestion, which can be used to perform model comparison under varying prediction conditions including outliers/anomalies and which can serve as a benchmark for future predictive work. 
We apply our models on data coming from counting stations along a long motorway in Australia, and we consider both the spatial structure of the datasets, as well as the historical records during a one year long dataset. 
We first present the data profiling, outlier detection and anomaly repair method, followed by the methodology for constructing various deep learning models such as CNN, LSTM and the hybrid architecture CNN-LSTM; we then compare them against typically employed approaches, such as ARIMA, individual station regressors, and the average historical traffic profiles.
We follow with a sensitivity analysis of historical versus prediction time horizons, and their impact on model performance.
We end with the comparative analysis of the prediction performances of the models under both outlier or regular data streams arriving, and conclude with a model choice discussion.

 }

\begin{figure*}[pos=t!,width=13cm,align=\centering]

	\centering
	\captionsetup{justification=centering}
		\includegraphics[width=0.99\textwidth]{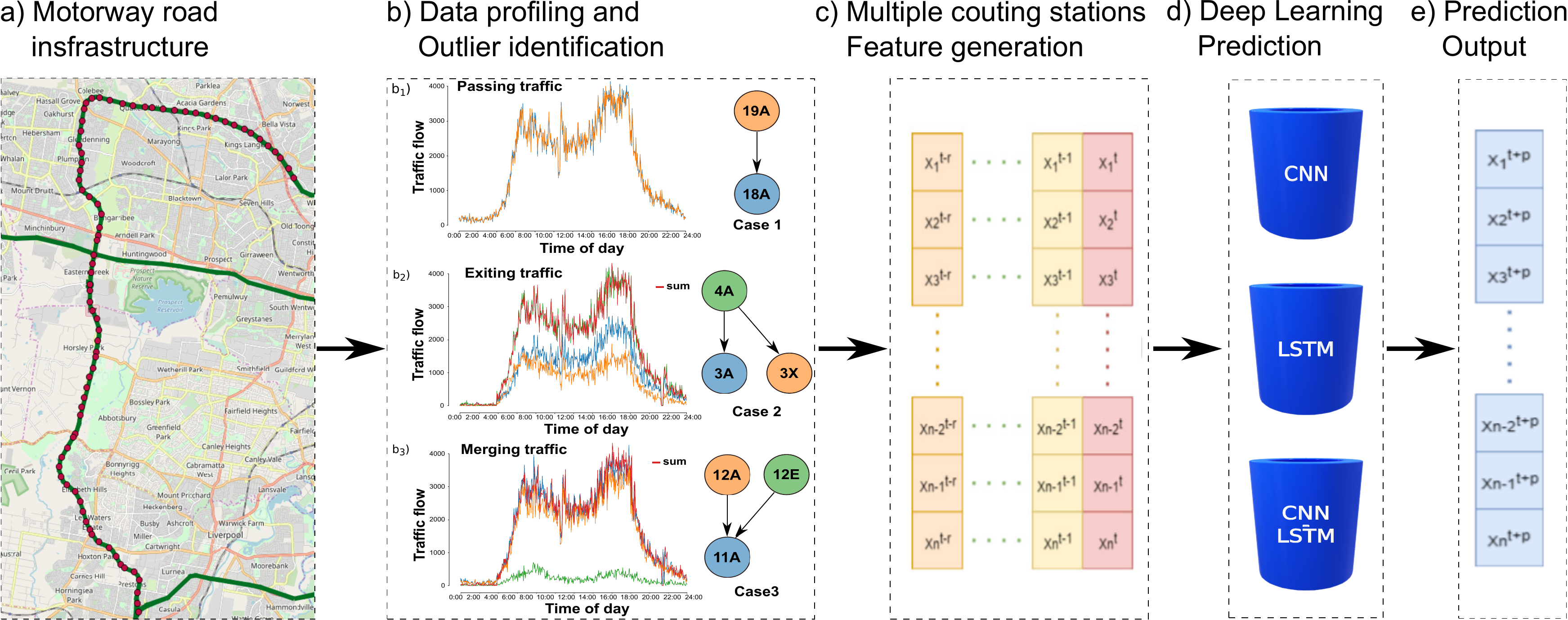}
	\caption{Schema of the proposed DL methodology for the motorway flow prediction.}
	\label{Fig_0_Methodology}
	\vspace{-0.5cm}
\end{figure*}

\section{Methodology}\label{S2_Method}

In this section we present the proposed deep learning methodology for predicting the traffic congestion along motorways.
\cref{Fig_0_Methodology} presents the proposed methodological framework, which consists of four steps: network identification (detailed in \cref{subsec:network-ident}), data profiling (\cref{subsec:profiling}), feature generation (\cref{subsec:feature-generation}) and DL model development and traffic prediction (\cref{subsec:models}).

\subsection{\textbf{Network identification and data set preparation}} 
\label{subsec:network-ident}

We first gather the spatial information with regards to the placement of traffic monitoring stations along the motorway segments, the road network geometry file and the temporal information in the form of traffic flow recorded at each time step (3min time-intervals in our case).
The obtained dataset is described in \cref{subsec:dataset}.

\subsection{ \textbf{Data profiling and outlier identification}} 
\label{subsec:profiling}
This step is necessary for building regular traffic patterns depending on the type of day, time-of-day, etc. 
\cref{Fig_0_Methodology} showcases 3 different possible cases of station spatial structure,
which are automatically checked for data accuracy and motorway structure consistency in both time and space. 
Let $T_f(\texttt{18A})$ and $T_f(\texttt{19A})$ be the traffic flows registered at the stations $\texttt{18A}$, and $\texttt{19A}$ respectively. 
If no exit or entry is recorded between two consecutive stations (see \cref{Fig_0_Methodology}-$b_1$), we assume that the two flow patterns should match, and we check for consistency as  $T_f(\texttt{18A}) = T_f(\texttt{19A}) \pm \epsilon$ ($\epsilon$ accounts for the inherent detector equipment error). 
When exit loops exist (see \cref{Fig_0_Methodology}-$b_2$), the module verifies that the sum of the flow recorded at the downstream stations ($\texttt{3A}$ and $\texttt{3X}$) matches the traffic flow recorded at the closest upstream station: $T_f(\texttt{4A})=T_f(\texttt{3A})+T_f(\texttt{3X}) \pm \epsilon$. 
Lastly, in case of entry loops (see \cref{Fig_0_Methodology}-$b_3$), the module checks that the sum of downstream traffic flow is the sum of upstream flows: $T_f(\texttt{11A})=T_f(\texttt{12A})+T_f(\texttt{12E})\pm \epsilon$. 
The data processing step also builds the daily flow patterns for all the stations along the entire motorway.
These are further used in the deep learning methodology, and to identify missing data and abnormal traffic flow due to traffic disruptions. 
This is further discussed in \cref{3_Case_study} together with speed and occupancy profiling as well. 
 
\subsection{\textbf{Feature construction}}
\label{subsec:feature-generation}

The traffic flow is recorded as time series associated with each monitoring station (including entries and exits). \verify{For this exemplification, we follow the discussion mainly around traffic flow, as similar approach is taken for speed and occupancy. }
It is processed and transformed into sequential matrices, which we denote as $\boldsymbol{X^t}$ and which are the input of our DL models: 
\begin{equation}\label{Eq1}
\boldsymbol{X^t}= 
\begin{bmatrix}
\boldsymbol{\vec{X_1}^t} \\
\boldsymbol{\vec{X_2}^t}\\
...\\
\boldsymbol{\vec{X_N}^t}
\end{bmatrix}
=
\begin{bmatrix}
x_1^{t-R+1}  & ... & x_1^{t-1}   & x_1^t \\
x_2^{t-R+1}  &... & x_2^{t-1}    & x_2^t \\
... 		 & ...        &...  & ...\\
x_N^{t-R+1}  & ... & x_N^{t-1}   & x_N^t \\
\end{bmatrix}
\end{equation}
where $N$ is the total number of monitoring stations along the motorway;
$x_j^t, j=\{1,...N\}$ is the traffic flow registered at station $j$ at the time point $t$;
$R$ is the number of historical points used to train the models; 
$t-R$ will be often referred to as the training horizon or a ``past time-window'' and $\boldsymbol{\vec{X_i}^t}$ the past horizon (training) vector for each station. 
 
Our prediction target is $\boldsymbol{\hat{X}^{t+P}} = [\hat{X}_{1}^{t+P}, \hat{X}_{2}^{t+P},... ,\hat{X}_{n}^{t+P}]^T$ where $P$ denotes the ``prediction horizon'' (how far in the future we want to make the prediction) and $\hat{X}_{j}^{t+P} = \begin{bmatrix} x_j^{t+1} & x_j^{t+2}  & ..&  x_j^{t+P} \end{bmatrix} $ is the predicted traffic flow at the $j^{th}$ station over the prediction horizon. A summary of all the notations used in this paper is provided in \cref{tab:notations}.
\begin{table}[pos=btp,width=8cm,align=\centering]

	\caption{Summary of notations.}
	\small
	\centering
	\begin{tabular}{cp{6.5cm}}
	\toprule
		Notation & Interpretation \\ %
		\midrule
		
		$N$ & the total number of stations used in this study ($N = 208$ comprised of $104$ in each direction). \\
		\ti & a 3-min \emph{Time Interval}; the time is discretized into 3-minute time intervals (480 \ti per day). \\
		$R$ & the length of the time window in the past; the number of \ti used as historic information.\\
		$P$ & future prediction horizon; predictions will be made for the $P^{th}$ \ti in the future. \\
		$x^{t-i}_j$ & the traffic flow of station $j$ at $\ti=t-i, i\in\{0,...R\}$.\\
				
		$\boldsymbol{\vec{S}^{t-i}}$  & the traffic flow for ALL stations at $\ti=t-i,i\in\{0,...R\}$; an $N$-dimensional column vector $\vec{S}^i$ = $[x^{t-i}_1$, $x^{t-i}_2$, ..., $x^{t-i}_N]^T$. \\
		$\boldsymbol{X^t}$ & the observed traffic flow, for all stations for the past $R$ \ti, an $R \times N$ matrix (see \cref{Eq1}). \\

		$RMSE$ & the Root Mean Square Error evaluation metric; $RMSE = \sqrt{\frac{1}{N} \sum_{j=1}^{N}\left(f_j-\hat{f_j} \right)^2 }$ \\
		$ReLU(x)$ & The ReLU function; $ReLU(x) = max(x,0)$ \\ \bottomrule
	\end{tabular}
	\label{tab:notations}
\end{table}

\subsection{\textbf{Deep learning model development}}
\label{subsec:models}

Deep learning is usually used to learn high dimensional functions via sequences of semi-affine non-linear transformations, and it has been shown effective general function learners~\cite{POLSON20171}.
A deep learning predictor is capable of addressing the non-linearity in the datasets, and of finding the spatio-temporal relations between features. 
We implement various DL models in the current modelling framework and we apply them either individually -- Convolutional Neural Networks (CNN), Long Short-Term memory networks (LSTM), back-propagation neuronal networks (BPNN) -- or in hybrid structures as an advanced deep learning architecture. 
We compare their performance to other parametric and baseline models in \cref{subsec:performance}. 
In the following we \revA{detail each of the DL} models and \revA{we} provide \revA{their} internal architecture setup. 

\paragraph{\textbf{Back-propagation Neuronal Networks (BPNN):}}

BPNN is a typical feed-forward type network which learns the relation between inputs and outputs without an explicit mapping of the information, and using a gradient descent optimisation method. 
Such models have been successfully applied for highway traffic incident detection \cite{CHENG2010482}, and also for tourist volume forecasting in Baidu \cite{LI2018116}.

The topology of BPNN usually includes an input layer, one or multiple hidden layers, and an output layer.   
In our DL work, we developed a BPNN model which consists of two fully-connected layers. 
The input of first layer is the historical information of all stations, and the last layer's output is the prediction of the traffic flow across all monitoring stations. 
In this work, BPNN is mainly used as a lower-bound DL performance measure, and it serves to assess the performance gains obtained when implementing the more complex models detailed here below.

\begin{figure}[pos=htb,width=8cm,align=\centering]
	\centering
	\includegraphics[scale=.65]{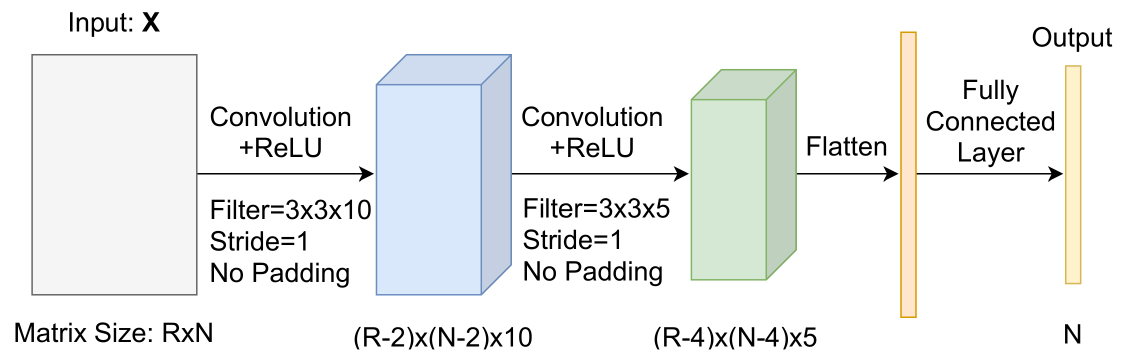}
	\captionsetup{justification=centering}
		\caption{CNN model for traffic flow prediction.}
	\label{Fig2_CNN}
	\vspace{-0.5cm}
\end{figure}

\paragraph{\textbf{Convolutional Neural Networks (CNN):}}
In order to take into consideration the spatial features of the traffic flow data, in which the individual counting stations reflect the propagation of flow in, out and along the motorway (see \cref{Fig_0_Methodology}-b)), we employ a CNN model on the traffic dataset.
CNNs are bio-inspired models which have been widely applied for processing images, speech and time series \cite{AAAI1714501}.
The main feature of CNN is the convolution operator, which slides on a two-dimensional surface, smoothing it and extracting higher level abstraction.
In image processing, multiple convolution operations are subsequently applied to increase the abstraction of the information.
The input of a CNN can include both a spatial dimension and a temporal dimension, and the convolution operator can be applied in either one dimension (1D) or both dimensions (2D).
In various advanced DL architectures, CNN models usually contain various convolutional layers with non-linear activation functions applied in between.

For our study, we construct a fully-connected CNN structure as presented in \cref{Fig2_CNN}, which accepts as input the RxN feature matrix $\boldsymbol{X^t}$ defined in \cref{Eq1}. 
This 2-dimensional input is passed through two convolutional layers and two Rectified linear activation unit (ReLU) functions, before finally being flattened as a 1-D vector and sent through a Fully Connected Layer which outputs the final results. 
No pooling layers are employed in our model; although, in general, pooling layers may increase the speed of training and achieve better performance on high dimensional image recognition tasks, the spatial dimension of our data set is smaller and the information it contains is not redundant like in images. 
For advanced DL architectures, the ReLU function are more popular than the traditional sigmoid/tanh functions because: a) they are more computationally efficient, b) they can help avoid the exploding and vanishing gradient problems and c) they tend to show better performance in practice \cite{NIPS2012_4824}.  

\begin{figure}[pos=h!,width=8cm,align=\centering]
	\centering
	\includegraphics[width=0.65\textwidth]{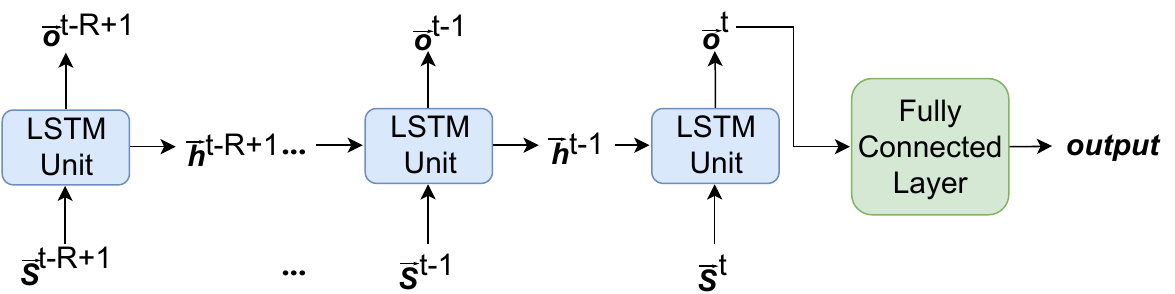}%
	\caption{LSTM model for traffic flow prediction.}
	\label{Fig3_LSTM}
	\vspace{-0.5cm}
\end{figure}

\paragraph{\textbf{Long Short-Term Memory Networks (LSTM):}}
The temporal features of the traffic flow have a different representation than  the spatial one.
While the traffic flow in one station can be localy determined by the neighbouring stations (see the three interconnected cases depicted in \cref{Fig_0_Methodology}-b)), various external events (e.g., accidents or weather conditions) can cause traffic congestion in the downstream part of the motorway, which propagates to the upstream stations and eventually affects the travel time along the entire motorway. 
In order to model these long-term dependencies in traffic flow, the long short-term memory units (LSTM) have been introduced for achieving a balance between immediate (short term) inputs and historical (long term) trends %

\cref{Fig3_LSTM} showcases the structure of the LSTM model that we develop in this work for the traffic flow prediction.
An LSTM \emph{unit} is typically comprised of an input layer, a hidden layer (which acts as a memory block containing an input gate, a forget gate and an output gate) and an output layer. 
The LSTM \emph{model} is a sequence of LSTM units, in which the output of one unit is consumed as input by the following unit.
The output of the last unit feeds into a fully connected layer which makes the the final prediction. 

In our application, the input feature matrix $\boldsymbol{X^t}$ is split into various flow vectors, one for each counting station (denoted as $\boldsymbol{\vec{S}^{t-i}}$). 
For each time step from the past horizon $\{t-R+1,.. t-1, t\}$, an LSTM unit accepts the vector $\boldsymbol{\vec{S}^{t-i}}$ as input and outputs a hidden state vector $h^i$ and an output vector $o^i$, of equal lengths. 
The hidden state $h^i$ is passed at the next unit $\{t-i+1\}$. 
The last output vector $o^t$ is connected to the fully-connected layer to get the final result. 
For our work, we only used only one hidden layer per unit and no drop-out layers, as we have a limited number of counting stations. 
For a more complex road network, this LSTM structure can be further extended.

\begin{figure}[pos=h!,width=9cm,align=\centering]
	\centering
	\includegraphics[width=0.65\textwidth]{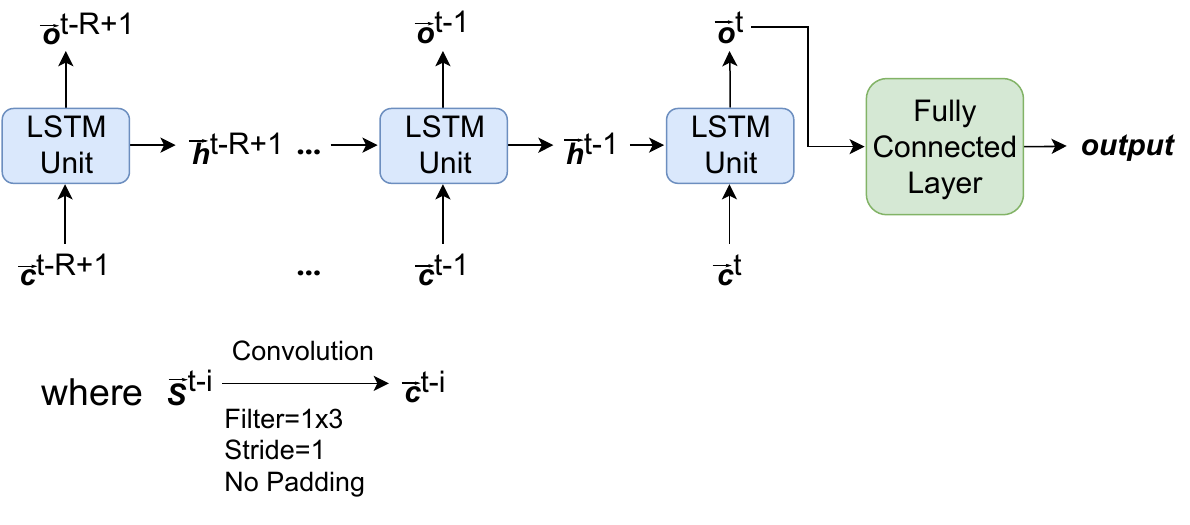}%
	\caption{CNN-LSTM hybrid model for traffic flow prediction. }
	\label{Fig4_CNN_LSTM}
	\vspace{-0.5cm}
\end{figure}

\paragraph{\textbf{Hybrid CNN-LSTM prediction:}}

We develop the hybrid model as
a combination of spatial and temporal processing, modelled by 
connecting the output of CNN to the input of each LSTM unit.
The intuition is that the structure of LSTM would learn the temporal patterns, while the structure of CNN can learn the location features. 
The final prediction is made using a Fully Connected layer, just like for LSTM. 
Several previous research works have employed the hybrid model and they report contradictory results: some argue that it improves the prediction accuracy, while others indicate the contrary~\cite{Nguyen2018}. 
\cref{Fig4_CNN_LSTM} showcases the structure of our hybrid model; the difference from a regular LSTM is that we use a 1-dimensional filter to scan the input (each $\boldsymbol{\vec{S}^{t-i}}$ is  processed by a 1-D convolution layer before it is fed into an LSTM unit). 
The added convolution layer has a 1x3 filter.
The results of each proposed model have been compared to other basic or parametric prediction models further detailed in \cref{S4_Exp}.

\section{Case study}\label{3_Case_study}

\subsection{The Sydney motorway dataset}
\label{subsec:dataset}
\verify{
Our methodology has been applied to a motorway traffic dataset, which was collected over the entire year of 2017, by recording the traffic flow, speed and occupancy at each of the 208 bi-directional ``counting stations'' along the M7 Motorway in Sydney, Australia (shown in \cref{Fig_0_Methodology}). 
The M7 motorway runs on the West of Sydney and it is the main motorway connecting North and South Sydney.
There are 104 metering stations in each direction including entries and exits; stations ending in A denote south-bound traffic, stations ending in B denote north-bound traffic while stations ending in E and X denote entries and exits respectively. The dataset contains 36.34 million data points for flow only, where for example one data point is the flow recorded by one station during one time-interval of 3min, denoted as \ti.}

\begin{figure*}[pos=htbp,width=\textwidth,align=\centering]

	\centering
	\subfloat[]{
		\includegraphics[height=0.173\textheight]{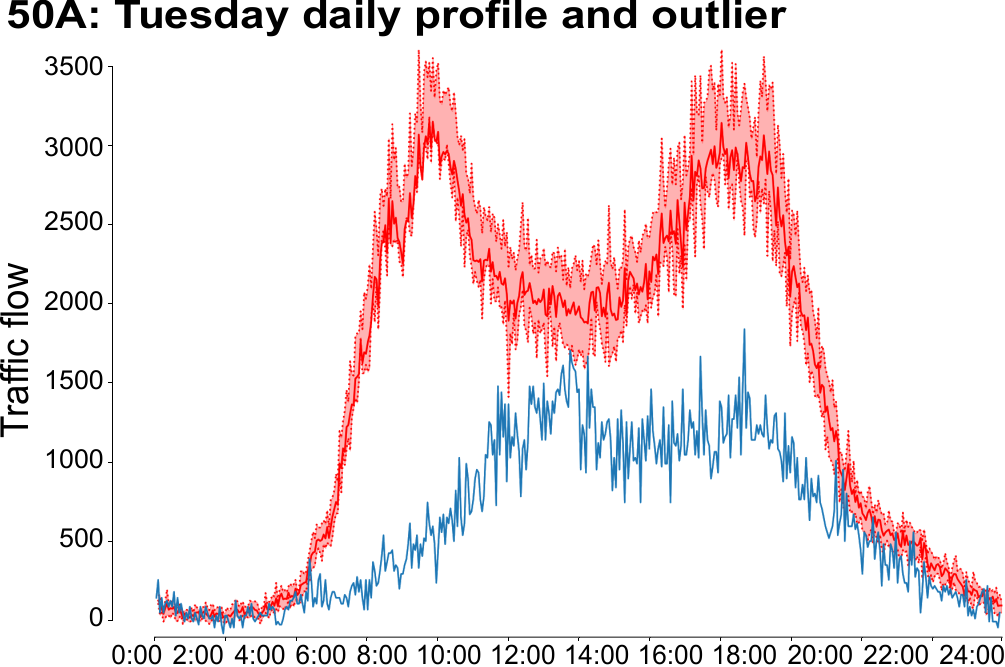}%
		\label{subfig:construct-daily-profile}%
	}%
		\subfloat[]{
		\includegraphics[height=0.173\textheight]{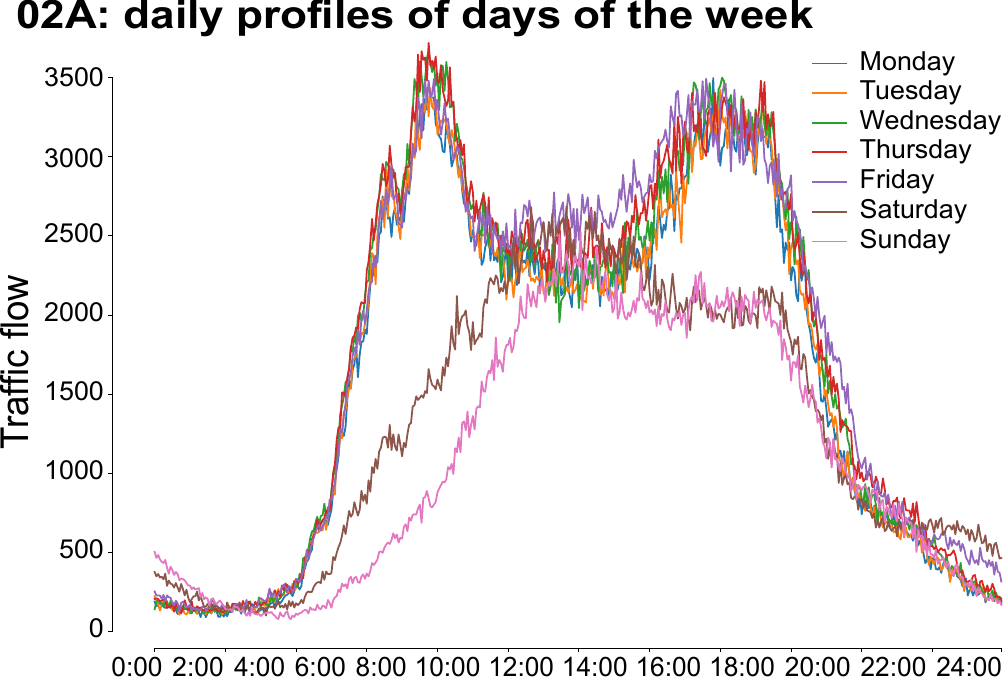}%
		\label{subfig:daily-profile-days-of-week}%
	}%
	\subfloat[]{
		\includegraphics[height=0.173\textheight]{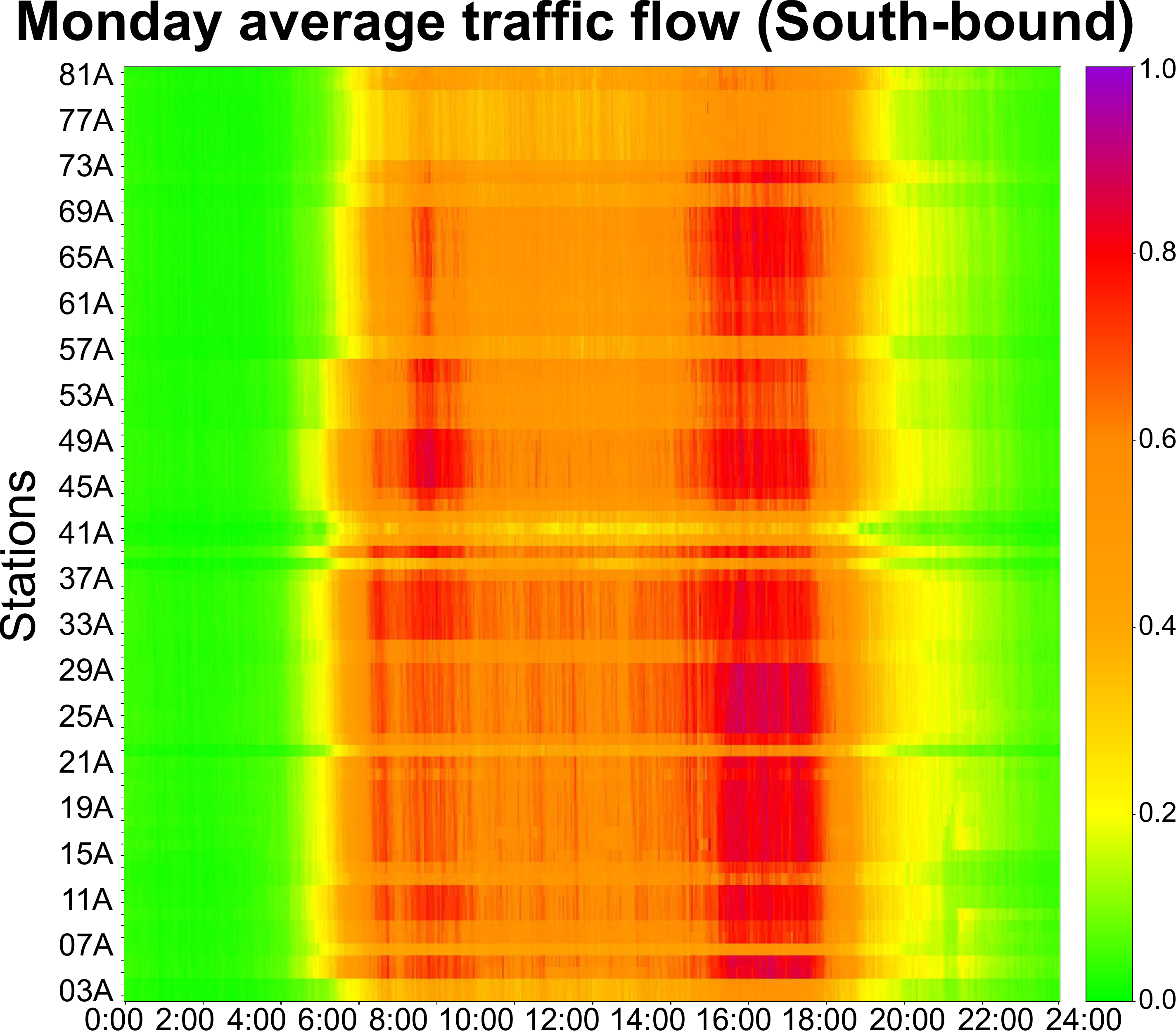}%
		\label{subfig:average-traffic-flow}%
	}%
		\centering

	\caption{
      	\textbf{(a) Constructing the daily profile.} 
      	Mean (solid line) and the $20\%-80\%$ percentiles (red area) for the traffic flow series for the station \texttt{50A}, computed on the period 2017-02-01 to 04-30;
      	\textbf{(b) Daily profiles for days of the week.} 
      	The daily profiles for station \texttt{02A} for each of the days of the week, computed for the same period of time.
      	\textbf{(c) Daily profiles for all stations -- the Traffic Flow Congestion Map.} 
      	The colormap of the Monday traffic flow for all 104 south-bound (A) stations is calculated based on a Flow/Capacity ratio and ranges between 0 and 1.
	}
	\vspace{-.3cm}
\end{figure*}

\subsection{Daily profiling}\label{subsec:Sec3_2_Daily_Profile}

\textbf{Daily profile.}
We start from the observation that the traffic flow at any given station presents a strong daily and weekly seasonality, mainly driven by the users daily work commute patterns.
We define the \emph{daily profile} as the typical daily traffic flow series recorded at a given station.
We compute a station's daily profile as the average flow for each \ti, for a given day of the week, over a period of three months (12 weeks in the period of February $1^{st}$ to April $30^{th}$ 2017).
\cref{subfig:construct-daily-profile} shows the mean flow for a Tuesday for the station \texttt{50A}, alongside with the $20^{th}$ and $80^{th}$ percentile values.
We make several observations.
First, the $20/80$ confidence interval wraps closely the daily profile, indicating that the 12 series are very similar and that the daily profile is a representative summarization.
Second, we observe that the daily profile shows two peaks, corresponding to the two rush hours: one in the morning (8-10 AM) corresponding to the daily commute towards work, and a second one in the afternoon (5-8 PM) corresponding to the end of the working day.
Last, the daily profile allows to identify non-standard days; for example, the blue line in \cref{subfig:construct-daily-profile} shows a significantly lower traffic, as it corresponds to April $25^{th}$ -- ANZAC day, a public holiday in Australia.

\textbf{Weekdays vs. weekend.}
\cref{subfig:daily-profile-days-of-week} plots the daily profiles for each of the days of the week, for station \texttt{02A}.
We observe that the weekdays (Monday to Friday) exhibit similar two peak patterns, whereas Saturday and Sunday have a single peak between 11AM and 4PM, and an overall lower flow.
Noteworthy, the ANZAC day flow shown earlier in \cref{subfig:construct-daily-profile} resembles a weekend pattern, despite being on a Tuesday.

\textbf{Traffic flow congestion map.}
\cref{subfig:average-traffic-flow} plots as a congestion map the Monday daily profiles of all South-bound stations by calculating the flow/capacity ratios. Here we also observe the two peak patterns of a typical weekday.
However, the traffic flow also allows to visually identify the most congested sections of the motorway (between \texttt{31A} and \texttt{13A} during the afternoon peak) and most importantly to track down abnormal congestion disruptions along the motorways. In addition to providing an abstraction of the typical flow, the daily profile is also used to correct the missing data in the dataset, most often occurring due to malfunctioning traffic recording devices for particular stations.

\begin{figure*}[pos=!htbp,width=\textwidth,align=\centering]
	\centering
		\includegraphics[width=0.8\textwidth]{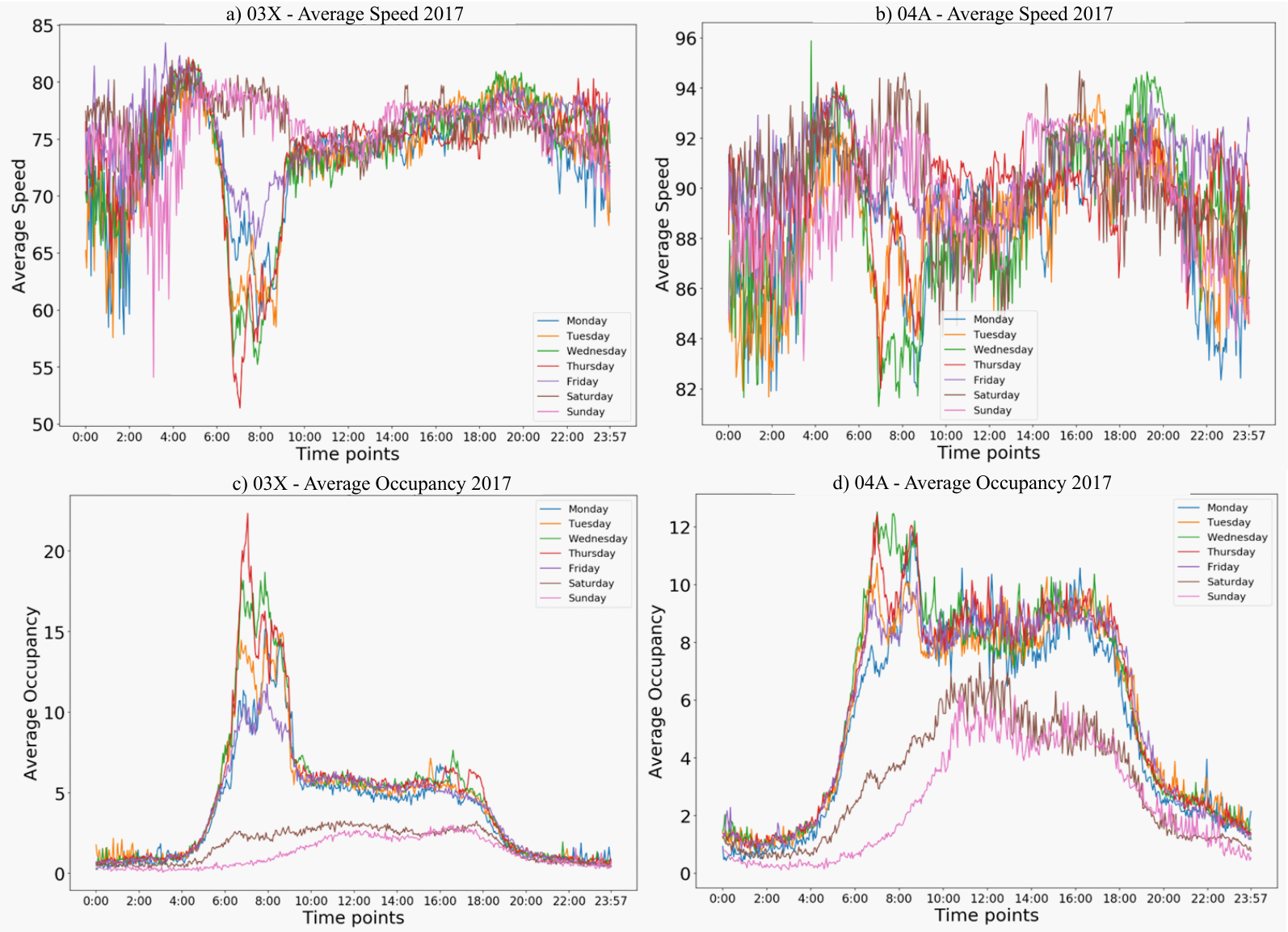}%
	\caption{ Speed daily profile for stations a) $03X$ and b) $04A$ respectively. Occupancy daily profiles for stations c) $03X$ and d) $04A$ respectively.
	}
	\label{subfig:Speed_occupancy_daily_profiles}%
	\vspace{-.3cm}
\end{figure*}

\verify{\textbf{Speed and occupancy daily profile:} Similarly to the traffic flow DPP, we build as well the daily profiles for Speed (SDPPs) and Occupancy (ODPPs), as represented in Fig. \ref{subfig:Speed_occupancy_daily_profiles}. We make the observation that the profile of a station can influence its DPPs as can been seen by comparing the exit station 03X and 04A (Fig. \ref{subfig:Speed_occupancy_daily_profiles}a) and b)):
while SDPPs of $04A$ has large variations and tends to be higher during the weekdays (Monday-to-Friday), the SDPPs of $03X$ seem to follow very similar trend at lower speed records, which has a significant speed drop during all morning peak periods between $06:30 AM$ and $09:30 AM$ to almost $55[km/h]$; this can reflect congestion for exiting the motorway into sub-arterial roads. Similarly, the ODPPs for $03X$ reflect higher peak during the same time as the speed drops in the morning rush hours of various weekdays, while for $04A$ the ODPPs score higher values for week days profiles, but very closely related to the SDPPs of the same stations. All DPPs for speed, occupancy and flow have been used for: baseline comparison of proposed DL models and anomaly detection and repairing procedure detailed in the following sections.      
}
\subsection{Missing data}

\begin{figure}[pos=h!,width=\textwidth,align=\centering]
	\centering
		\includegraphics[scale=0.18]{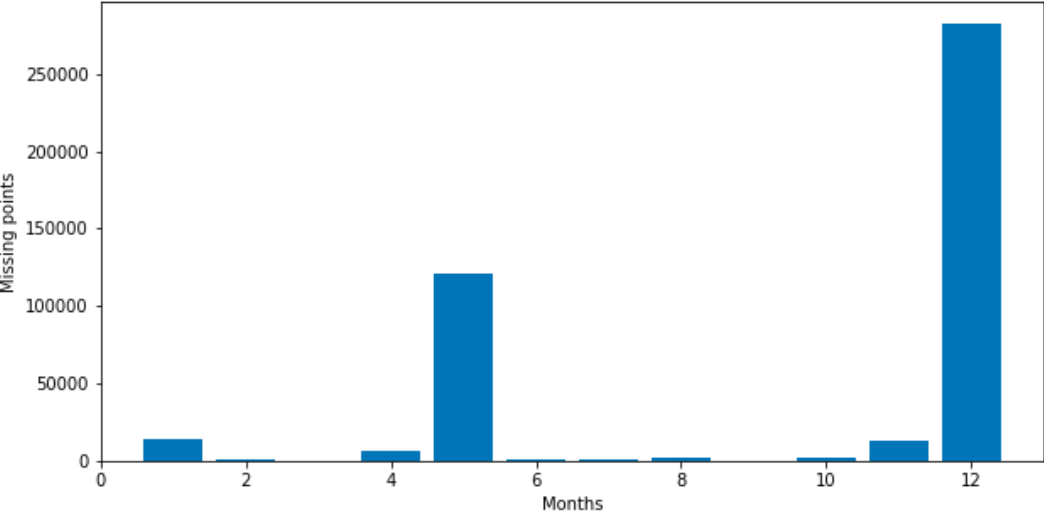}%
		\caption{The total number of missing data points in the Sydney Motorway traffic flow dataset, aggregated per month.}
		\label{si-subfig:missing-data-counts}%
\vspace{-0.5cm}
\end{figure}

\verify{
Upon initial inspection, the Sydney Motorway traffic dataset contains missing data points from the real data streaming which are misaligned with the overall traffic behavior in that area.
We detect such malfunctions at the level of the entire dataset, and we count them per month. Fig. 
\ref{si-subfig:missing-data-counts} shows that there is an abnormally high number of missing data in the months of May and December 2017.
In our initial work, we excluded these two months from our training data (as discussed in \cref{S4_Exp} of the main paper) and conducted our Deep Learning Training on a selected months form our data set. However, a real-time prediction engine should be applied on all type of data sets, and should be able to detect very fast anomalies and outliers that might be transmitted in real time. For this reason, we have extended our previous work and constructed an anomaly detection engine which follows several steps described in the next subsection. The final results will focus on presenting results under both types of set-ups: those using only the original data set without any anomaly detection and those using the outlier processing procedure making use of daily profiles and data interpolation presented in the following.

}

\subsection{Anomaly detection and data repairing}

\begin{figure}[pos=htb,width=9cm,align=\centering]
	\centering
            \includegraphics[width=1\columnwidth]{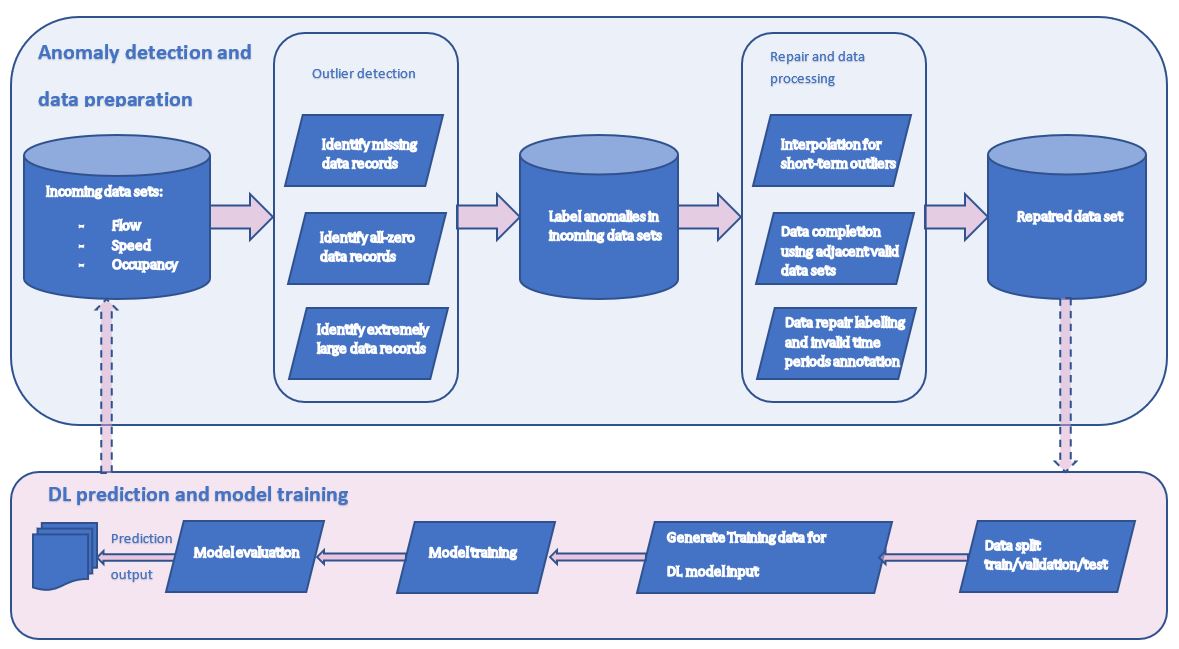}
	\caption{ \verify{Anomaly detection and repairing methodology.}}
	            \label{fig:Anomaly_detection_diagram}
	\vspace{-0.5cm}
\end{figure}

\verify{
Before developing the Deep Learning Methodology and its relevant models deployed for our study, we firstly apply an anomaly detection and data repairing procedure which is represented in \cref{fig:Anomaly_detection_diagram}. Based on the incoming data sets such as flow, speed and occupancy of the motorway, we apply several steps to: a)identify the outliers in the data (these can be of different types and can appear as either missing records, all-zero records or extremely large records), b) to treat the outliers by applying various steps such as interpolation, adjacent data completion based on motorway structure and fundamental diagram represenation of each station, and finally c) annotate the repaired data set and prepare the release for deep learning model training, testing and validation. We make the observation that once the DL models have produced the predicted results over a future selected time horizon, the procedure can repeat again based on new incoming data sets for each motorway station. The following sub-sections detail all steps and modules presented in \cref{fig:Anomaly_detection_diagram} by looking into the motivation of addressing each problem and the description of steps taken to address them.      
}        

\verify{

\begin{enumerate}
  \item \textbf{Identify missing data records}
    \begin{itemize}
            \item \textbf{Motivation}: 
        In the incoming dataset $\mathcal{D}_O$ is not complete, some time points might be missing from being transmitted. Such missing records may be the result of equipment maintenance or can arise from the network failure. Without addressing these missing points, the data set will not be continuous and hard to be used to train any deep learning models. Fig. \ref{fig:miss01} showcases an example of missing records from an entire 24-hour traffic flow profiling; the Ox axis represents the time interval (with a 3-min frequency) while the Oy axis represents the listing of traffic flow for each station from $02A$ until $72A$. The traffic flow is coloured by its values with highest values reaching 5,000 vehicle/3-min time interval. The dark region represents almost 3 hours of missing data points from the stream which have unidentified cause. 
        
        \item \textbf{Description}: 
        We assume that data stream  $\mathcal{D}_O$ should transmit all the records of all stations $s$ at all time points between a start time point $T_{start}$ and an end time point $T_{end}$ with a certain frequency (time-interval) $TI$, where $T_{start}$ is 00:00:00 01/01/17, $T_{end}$ is 23:57:00 31/12/17 and $TI=3$ minutes. The set of all possible records is denoted as $\mathcal{R}_{complete}$ and should contain all data points. For example, if there is no incoming flow record, we will label it $r_{miss}$ and add it into a sub-set of missing points $\mathcal{R}_{miss}$.
     \end{itemize}
        
        \begin{figure}[pos=htb,width=\textwidth,align=\centering]
            \centering
            \includegraphics[width=0.7\columnwidth]{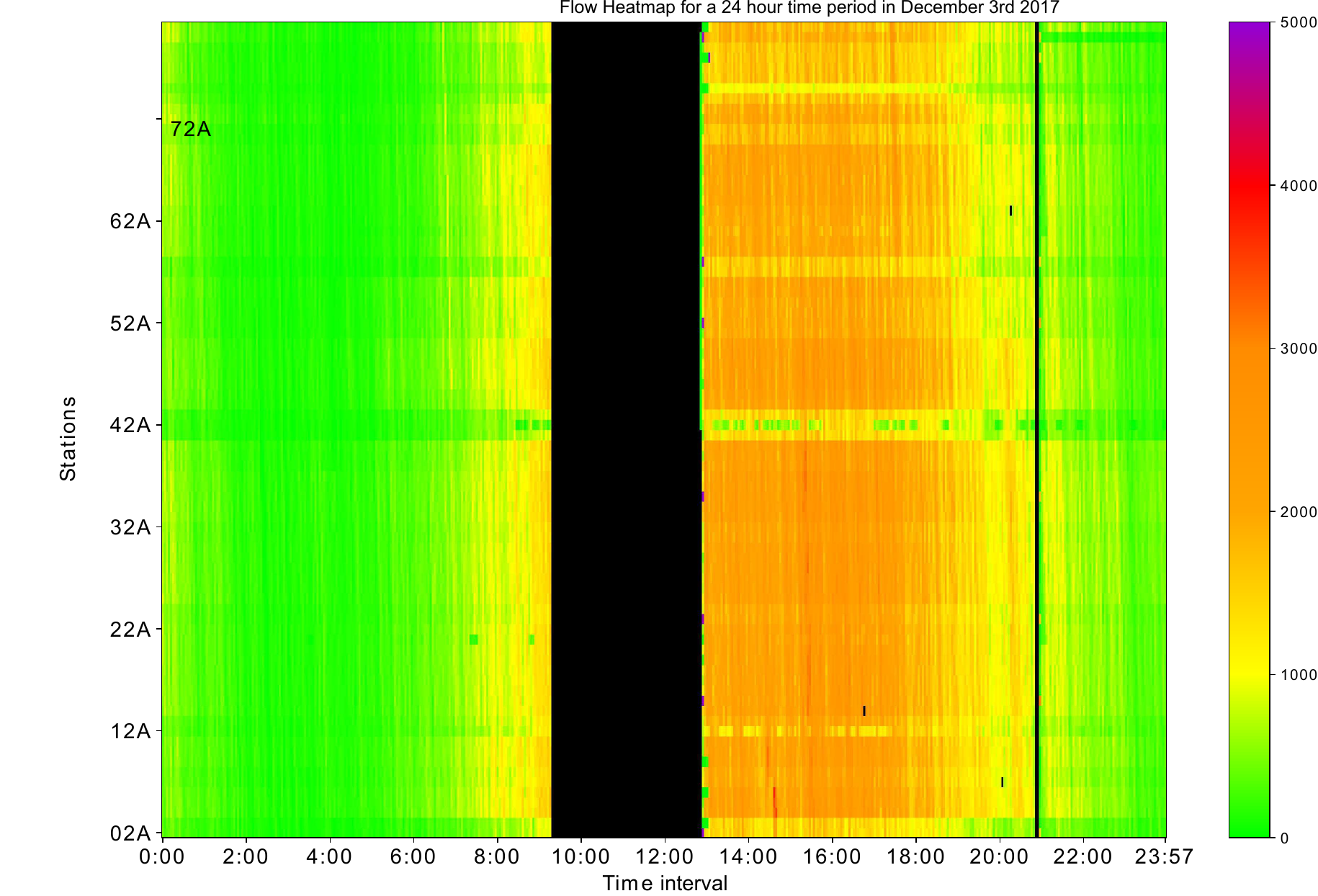}
            \caption{Example of missing records. The records in black area are not included in original dataset and may contain valuable information on how the congestion evolved during the missing time-period.}
            \label{fig:miss01}
            \vspace{-0.5cm}
        \end{figure}
               
  \item \textbf{Identify all-Zeros data records}
    \begin{itemize}
        \item \textbf{Motivation}: The initial data profiling showed for example that in some cases, a station's flow, speed and occupancy can suddenly drop to zero while its neighbourhood stations (preceding and following) continue to report high traffic flow counts for example. When no entries/exits are present in between consecutive traffic flow stations, then the data profiling records should all follow similar trends. If one of the consecutive stations does not align with its neighbours reporting data streams, this indicates another erroneous data transmission or device malfunction. %
        Similar to the missing records, we also want to locate all abnormal all-zeros records that might occur and repair/adjust them in following steps. Fig. \ref{fig:zero01} showcases an example of all-zeros records recorded by the station $23A$ which should follow the same traffic flow trends as its neighbours $22A$ and $24A$ as no entries/exit ramps are in between them. However the flow recorded by station "$23A$" seems to be suffering from various drop-outs throughout the day possibly due to device malfunctioning or wrong-reporting.

        \item \textbf{Description}: We make the observation that some traffic flow records which are all-zeros can appear during the night time as there is less traffic on the highway, and they should be considered as correct, which means the records of adjacent time points and stations also have low traffic flow. However, all of all-zeros records that can appear during daytime are abnormal, especially under non-zero neighbourhood reporting. To mitigate and differentiate between these difference occasions, we count all zero-records and store all records which have time timestamp between 8:00 AM to 9:00 PM into a different data set which we name $\mathcal{R}_{zero}$.
    \end{itemize}

    \begin{figure}[pos=htb,width=\textwidth,align=\centering]
	\centering
	\subfloat[]{%
            \includegraphics[width=0.5\columnwidth]{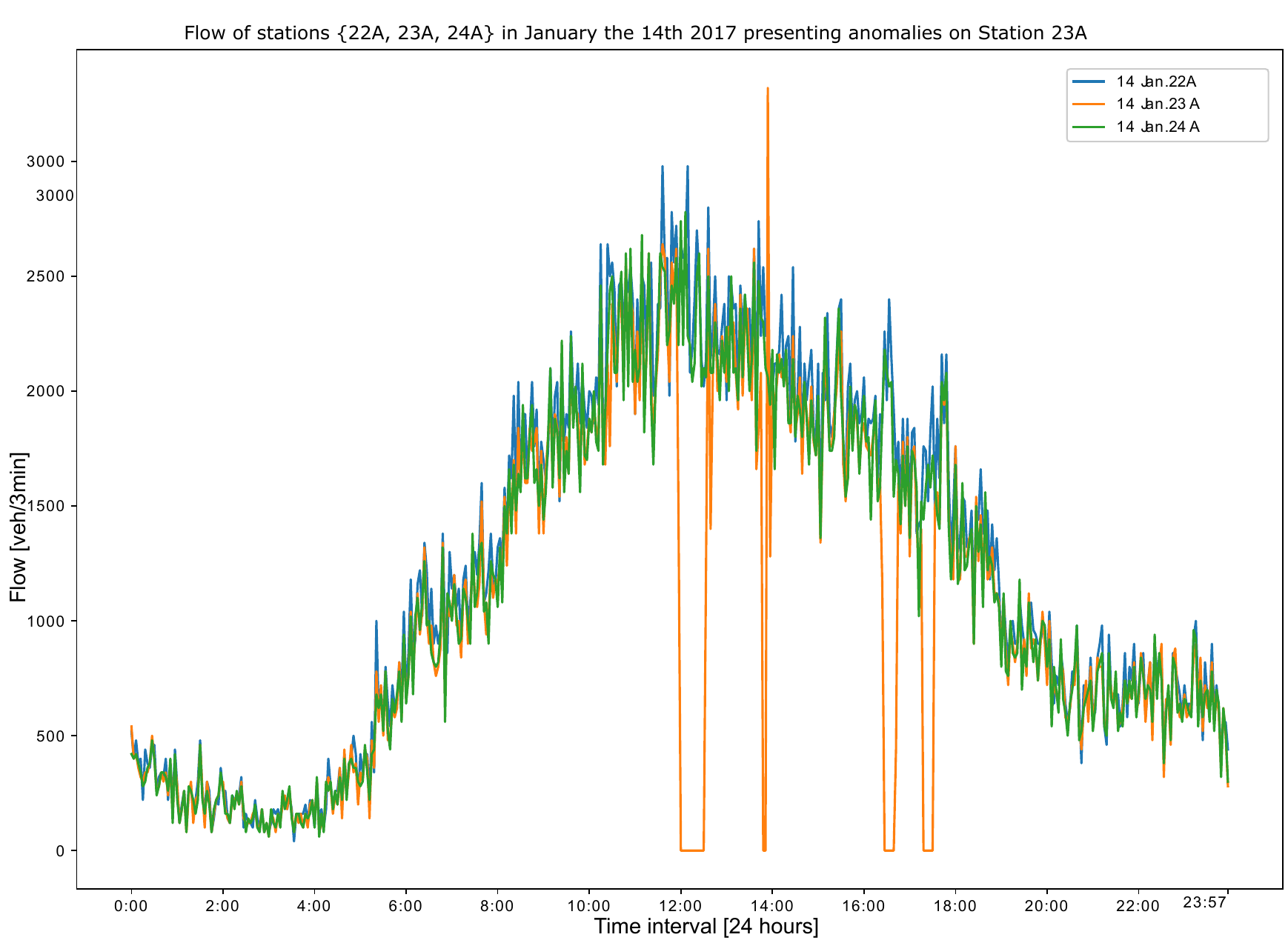}
            \label{fig:zero01}
	} 
	\subfloat[]{%
            \includegraphics[width=0.5\columnwidth]{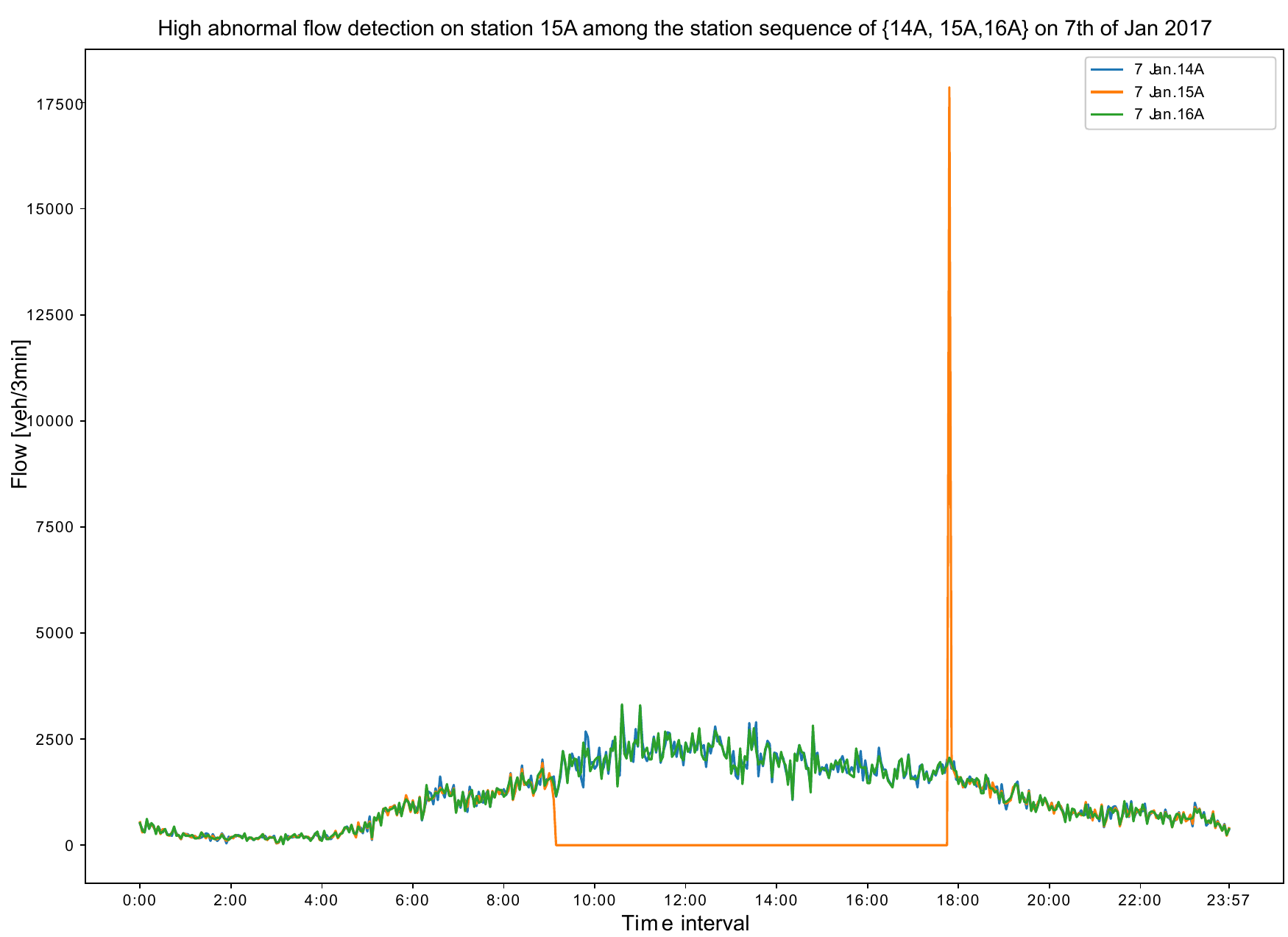}
            \label{fig:high01}
	}%
	\caption{ a) Example of abnormal all-zeros records. This figure shows the flow curve of three adjacent stations in same day. While the flows of previous station (24A) and next station (22A) are not changed significantly, the flow recorded by station 23A suddenly drop to zeros in several times. b) Example of high-flow records. This figure shows the flow curve of three adjacent stations in same day. While the flows of previous station (14A) and next station (16A) are as usual, the flow recorded by station 15A suddenly drop to zeros and last for a while. Then the flow sharply increases to a very high value in next time point.}.
	
	\vspace{-0.5cm}
	
\end{figure}

  \item \textbf{Repairing Long Abnormal All-Zeros Periods}
    \begin{itemize}
        \item \textbf{Motivation}: Another finding of the data profiling is that a few traffic count stations can present continuous abnormal all-zeros records which can last up to several days not only a few hours as described in the first step of this subsection. Such long abnormal all-zeros periods will heavily affect the correctness of the daily profiles and also the training of any deep learning models that we deploy. So we need to detect and repair this type of anomaly before conducting other data preprocessing or analytic investigations.
        
        \item \textbf{Description}: To mitigate this problem, firstly we compute the daily profiles of each station without considering the records in $[ \mathcal{R}_{miss};\cup\mathcal{R}_{zero} ]$. A daily profile trend (DPP) includes the mean, median and standard deviation of flow, occupancy and speed of every station in every time points of every reporting day. We use the historical DPP as a reference on how traffic flow "should" look like giving historical information when no new data sets are coming in. We then merge the adjacent abnormal all-zeros records as abnormal all-zeros periods. For the long periods which are longer than 2-hours, we replace their zero values by their corresponding daily profile values. The decision of the 2-hour threshold for missing values has been considered after looking into reported traffic accidents throughout the year; we therefore consider that any period with missing values longer than 2-hours reflects a serious defect of the station, and not just a data reporting malfunction. The short anomaly periods (less than 2 hours) are repaired in the last step of this procedure. The repaired dataset is noted as $\mathcal{D}_{R1}$. Noticing that the records in $[ \mathcal{R}_{miss};\cup\mathcal{R}_{zero} ]$ are excluded from $\mathcal{D}_{R1}$.
    \end{itemize}
        
  \item \textbf{Identify Extremely Large Data Records}
  
  \begin{figure}[pos=htb,width=\textwidth,align=\centering]
	\centering
            \includegraphics[width=1\columnwidth]{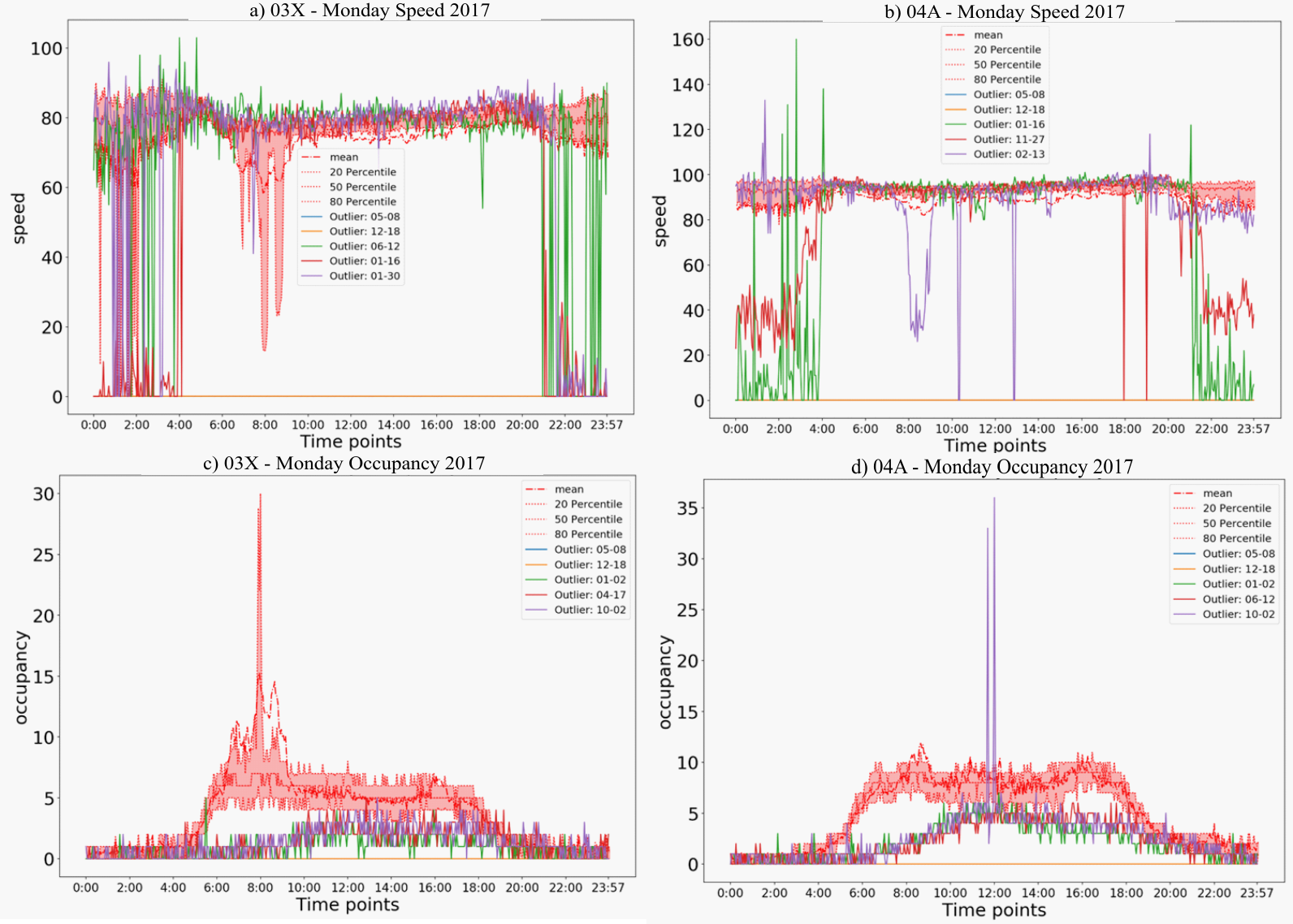}
	\caption{ a) Examples of abnormal all-zeros records for speed recorded on Monday for two different stations: a) $03X$ (exit) and b) $04A$ (regular). Similarly, some examples of abnormal and high accumulated records for occupancy for c) $03X$ (exit) and d) $04A$ (regular).}
	\label{fig:Speed_occupancy_outliers}
	\vspace{-0.5cm}
\end{figure}

    \begin{itemize}
        
        \item \textbf{Motivation}: Similarly to the situation of abnormal all-zeros records, some stations suddenly report very high values of traffic flows while their neighbourhood stations are following the usual historical DPP. As an observation, we noticed that this situation usually happens after a long abnormal all-zeros period of time. However, not every high-flow records can be matched to an all-zeros periods and some situations may reflect increased traffic circulation along the highway. Therefore we use daily profile to locate them in combination with the fundamental diagram which is further p
        
        Fig. \ref{fig:high01} showcases an example of high-flow values being recorded by an intermediary flow count station which should following similar trends with its neighbours. Once the station re-starts reporting the traffic flow, its values reach 17,500 veh/3min TI compared to maximum 2,500 veh/3min TI which were reported by its neighbours. 
        
        We also make the observation that both cases presented previously can appear under different set-ups based also on the profile of the counting station (whether is an entry/exit or main station). Two other examples are provided in Fig. \ref{fig:Speed_occupancy_outliers} for speed and occupancy anomaly detection against DPP for an exit station ($03X$) and a regular close-by station ($04A$); more specifically, Fig.\ref{fig:Speed_occupancy_outliers}a) and b) represents the speed DPP (mean, the $20$th and $80$th percentiles) versus several days presenting outliers in terms of "abnormal zero records" from the speed transmission. Station $04A$ showcases few ``abnormal'' all-zero records followed by very high records during night time (from $22:00$ - $04:00$) mostly on 27/11 and 16/01; it also presents a different abnormal behaviour on 13/02 when there is a sudden speed drop around 8:00 AM followed by 2 all-zero record transmissions around 10:15 and 12:30. This situation can reveal a reported accident around the morning peak followed by a slowed traffic movement around the highway resulting in an accumulation of reported vehicles as seen from the occupancy of the station showcased in Fig. \ref{fig:Speed_occupancy_outliers}d) 30 minutes prior to the speed drop from 12:30. Station $03X$ has a different profiling and more abnormal speed drops during the night time mostly due to the fact that less vehicles take this exit outside daylight hours. However, this exit station presents a very low $20$th percentile around the 8AM peak hours when the speed is dropping to the limit of almost 20km/h while the occupancy increases sharply out to 30 vehicles/TI. This is a reflection of many vehicles taking the exit for ``work'' commuting which causes a significant speed drop. This situation should not be treated as an anomalous event but rather as a congested traffic behaviour for morning rush hours. 
        
       The above analysis made space for using as well the Speed-Flow Diagram representation of each station when searching and identifying the anomalous events and adding extra verification conditions. Fig. \ref{Speed_flow_diagram} showcases the Speed-Flow diagram from a main station $16A$ on Mondays, by splitting the traffic data records in two: peak versus non-peak transmissions. The maximum recorded flow for this station is around 4850 veh/15 min time-interval recorded at a speed of almost 87km/h (S2). The area can be divided in several regions and speed-flow thresholds which we use as well in the anomaly detection verification procedure. For example, while area $A1$ indicates high operating speeds across the motorway under lower flow conditions (low non-peak traffic behaviour), area $A2$ reveals high traffic under peak hours operating at still very good speed limits. On the contrary, if the number of vehicles on the motorway is very high but speed start dropping, this reveals high congestion accumulating on the highway, with bottlenecks and potential disruptions that can appear and longer travel times. However, when lower flow is detected at lower running speeds (see area $A3$) this can reveal a potential reported incident (coupled as well with high occupancy); lastly but not least, area $A5$ is the one reflecting very low thresholds in both speed and flow, which can indicate either data anomaly, outliers or abnormal behaviour; this coupled with very low occupancy will reinforce the anomaly detection algorithm. The detected thresholds are used for conditional verification among all counting stations and their historical DPP versus new incoming data streams.         
        
          \begin{figure}[pos=htb,width=\textwidth,align=\centering]
	\centering
            \includegraphics[width=0.6\columnwidth]{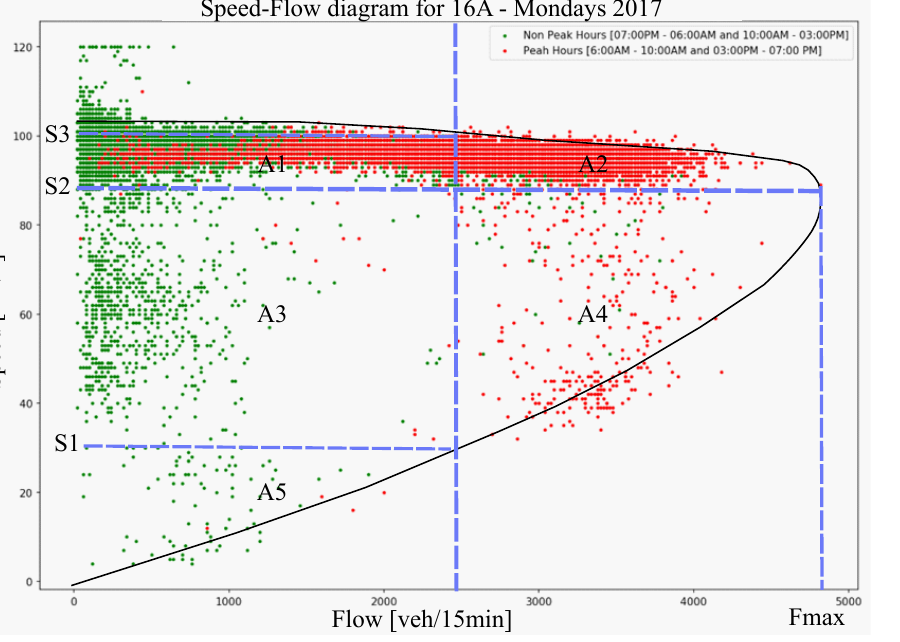}
	\caption{ \verify{Speed-Flow diagram for a selected station $16A$ showcasing various levels of congestion, maximal speed and regular operating speed and flow levels.}}
	            \label{Speed_flow_diagram}
	\vspace{-0.5cm}
\end{figure}

        \item \textbf{Description}: For each records in $\mathcal{D}_{R1}$, we firstly get its corresponding median and standard deviation from the computed daily profile. If the record flow is higher than the median by a large margin (which we consider to be as 10 times the standard deviation of the median traffic flow for that particular day of the week), we start verifying the traffic conditions according to the above verifications on speed-flow-occupancy; if all anomaly conditions are met, then we remove the record from $\mathcal{D}_{R1}$ and add it into a separate data set named $\mathcal{R}_{high}$. The resulted data set is noted as $\mathcal{D}_{R2}$. 
     If after detecting the a) traffic missing records, b) abnormal all-zeros records and c) high-flow records, we detect various anomaly combination with very high frequency throughout the data transmission extending over more than 2 hours at a time as well, we mark that day as ``unreliable'' and extract it from the DL model training. The invalid data record will now be  $[ \mathcal{R}_{miss}; \cup \mathcal{R}_{zero}; \cup \mathcal{R}_{high} ]$.
        
    \end{itemize}

  \item \textbf{Repairing anomalies and data processing}
  
  \begin{itemize}
        \item \textbf{Motivation}: The simplest method to repair invalid periods is to directly replace them by their corresponding daily profiles. However, as previously shown, the traffic condition is very complex and may be affected by many factors, including traffic accidents, weather, public events, etc. Therefore, while repairing the invalid or anomalous data records, we also need to take into account other valid records from the same day or from the neighbouring stations. This a two-step approach described below.  
        
        \item \textbf{Description}: We therefore propose an outlier repairing strategy which would fix invalid periods on a station by station approach and day-by-day by applying an interpolation followed by a repairing strategy to minimise errors between average DPPs and daily reported incoming data streams. For a certain station and a certain day, we denote all invalid time point as $T_{invalid}$ and all valid time point as $T_{valid}$. We also estimate the mean features (flow, occupancy and speed) of this station in this weekday from its DPPs. We start by keeping track of latest data points received by the station to have a tracking of the latest trend evolution of the data streams (keeping last 15 minutes reported data points) and matching against the DPPs and surrounding stations. If one or multiple of the above described anomaly situations is/are found, then we apply an adjustment (repairing) strategy by applying a minimisation technique (least square error) between the incoming data stream of latest valid records and the mean data streams (DPPs); this revolves at identifying the following two parameters $\alpha$ and $\beta$ expressed as: 
        
        \begin{equation}
        \alpha, \beta = argmin_{\alpha, \beta} \left( \sqrt{ \frac{ \sum\limits_{t \in T_{valid}} (F_t - \alpha\bar{F_t} - \beta)^2 }{length(T_{valid})} } \right)
        \end{equation}\label{Eq.1}

        where $F_t$ is the vector of valid records (can be flow, speed and occupancy) and $\overline{F_t}$ is the corresponding mean vector from the daily profile.
       Furthermore, according to the calculated $\alpha$ and $\beta$, we can adjust can now calculate the repaired flow ${F}^R$ as: 
        \begin{equation}
            {F}^R = \alpha\overline{F} + \beta
        \end{equation}
        where  $\overline{F}$ is corresponding mean vector of valid records. 
        
	\end{itemize}
	
\end{enumerate}
} %

\subsection{Anomaly detection and repairing results}

\begin{figure}[pos=htb,width=\textwidth,align=\centering]
	\centering
	\subfloat[]{%
		\includegraphics[width=0.5\columnwidth]{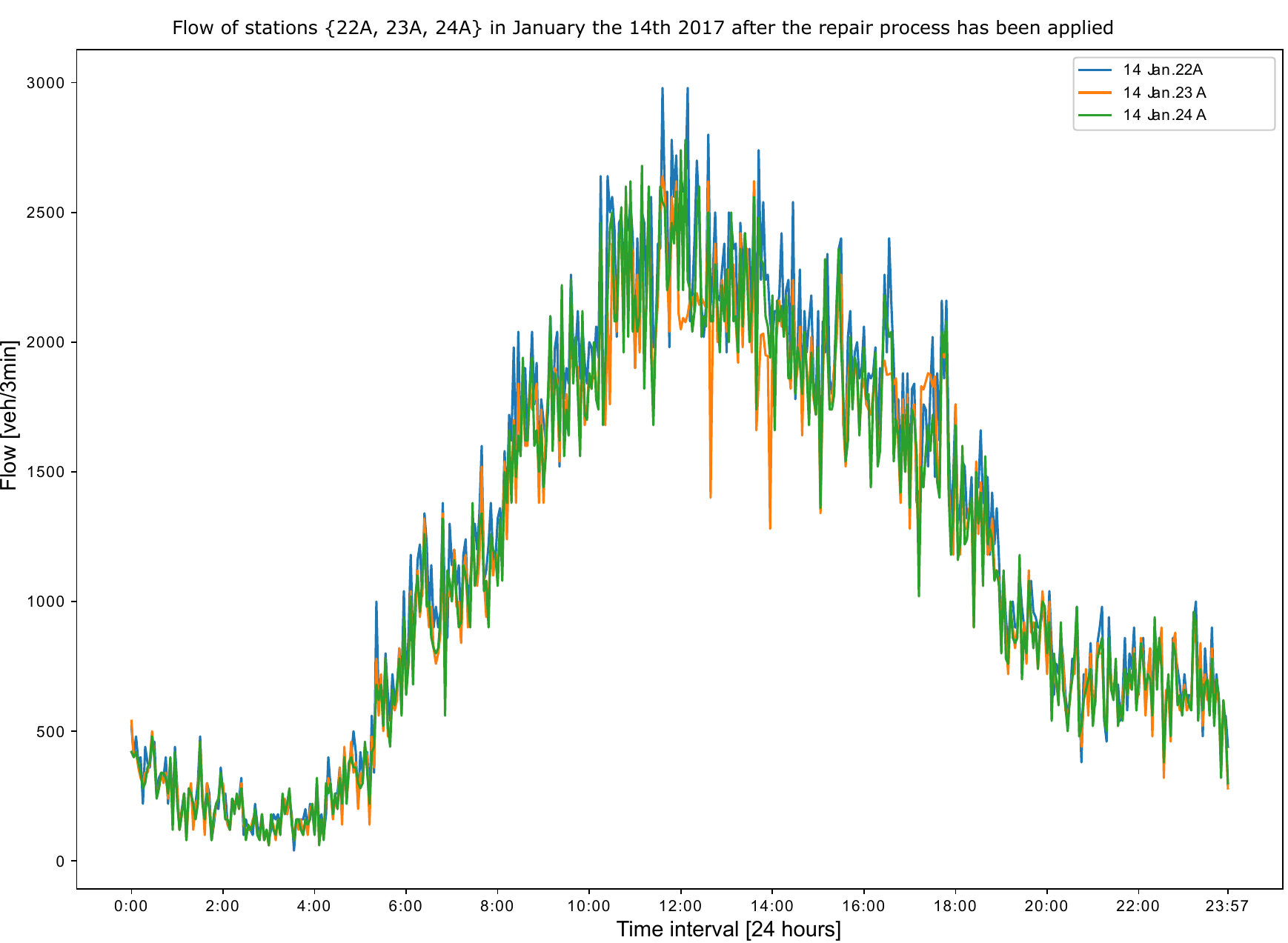}
            \label{fig:zeroRe01}
	} 
	\subfloat[]{%
		\includegraphics[width=0.5\columnwidth]{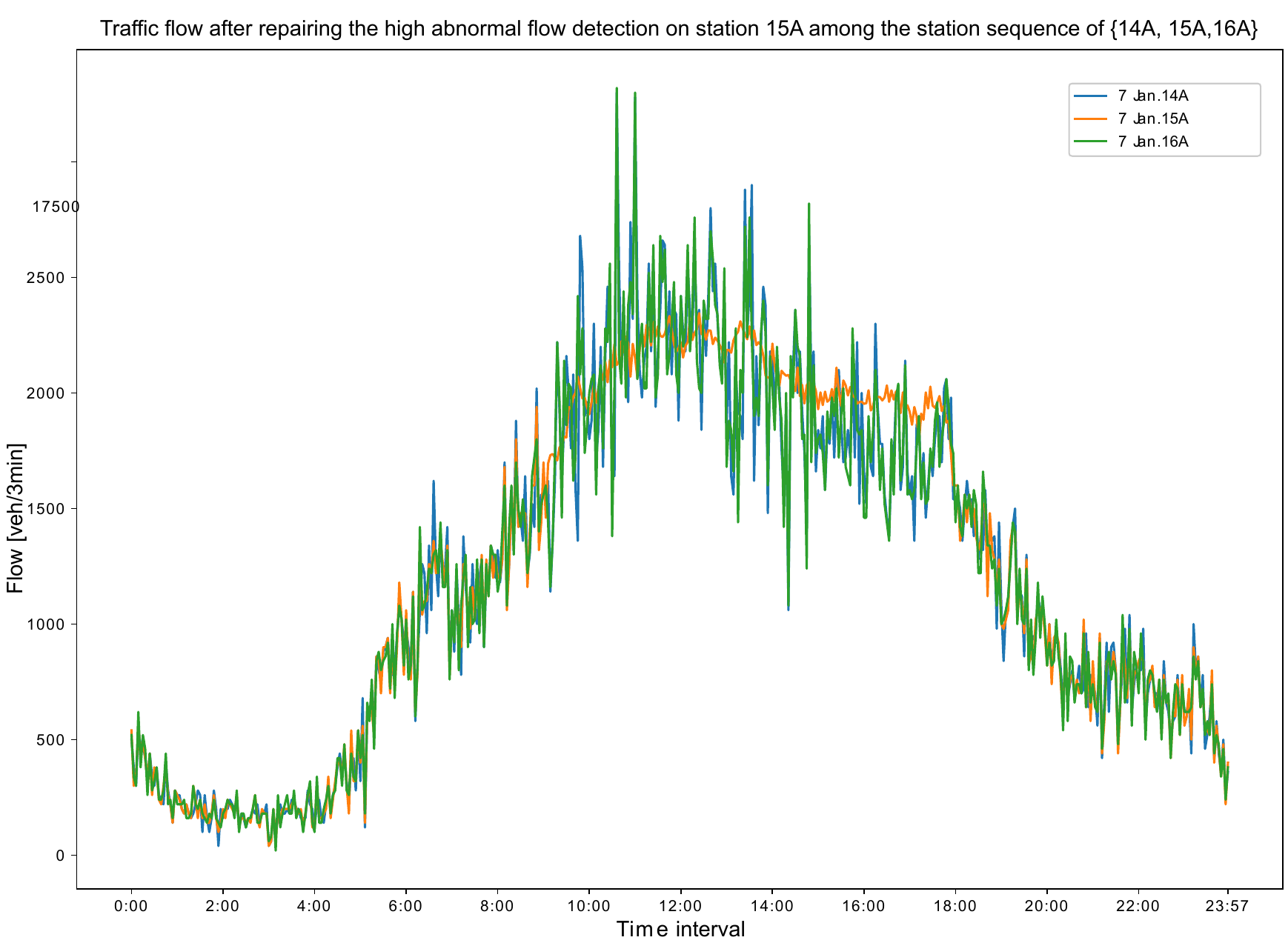}
            \label{fig:highRe01}
	}%
	\caption{ The repaired version of a) Fig. \ref{fig:zero01} and b) Fig. \ref{fig:high01}.
	}
	\vspace{-0.5cm}
\end{figure}

\verify{

Figures \ref{fig:zeroRe01} and \ref{fig:highRe01} represent the results obtained after applying the proposed method described in the steps above. More specifically, Fig. \ref{fig:zeroRe01} showcases how the traffic flow o station $23A$ looks like after the repairing of all zero missing values from the transmission, while Fig. \ref{fig:highRe01} represents the final results after repairing the flow of station $15A$ when high traffic flow records have been reported. We do make the observation that repairing high traffic volumes missing might lead to an increase in the repaired data set compared to it neighbours, however the traffic flow remains under the median and confidence interval calculated for this particular station and day. 

\begin{figure}[pos=htb,width=\textwidth,align=\centering]
	\centering
            \includegraphics[width=1\columnwidth]{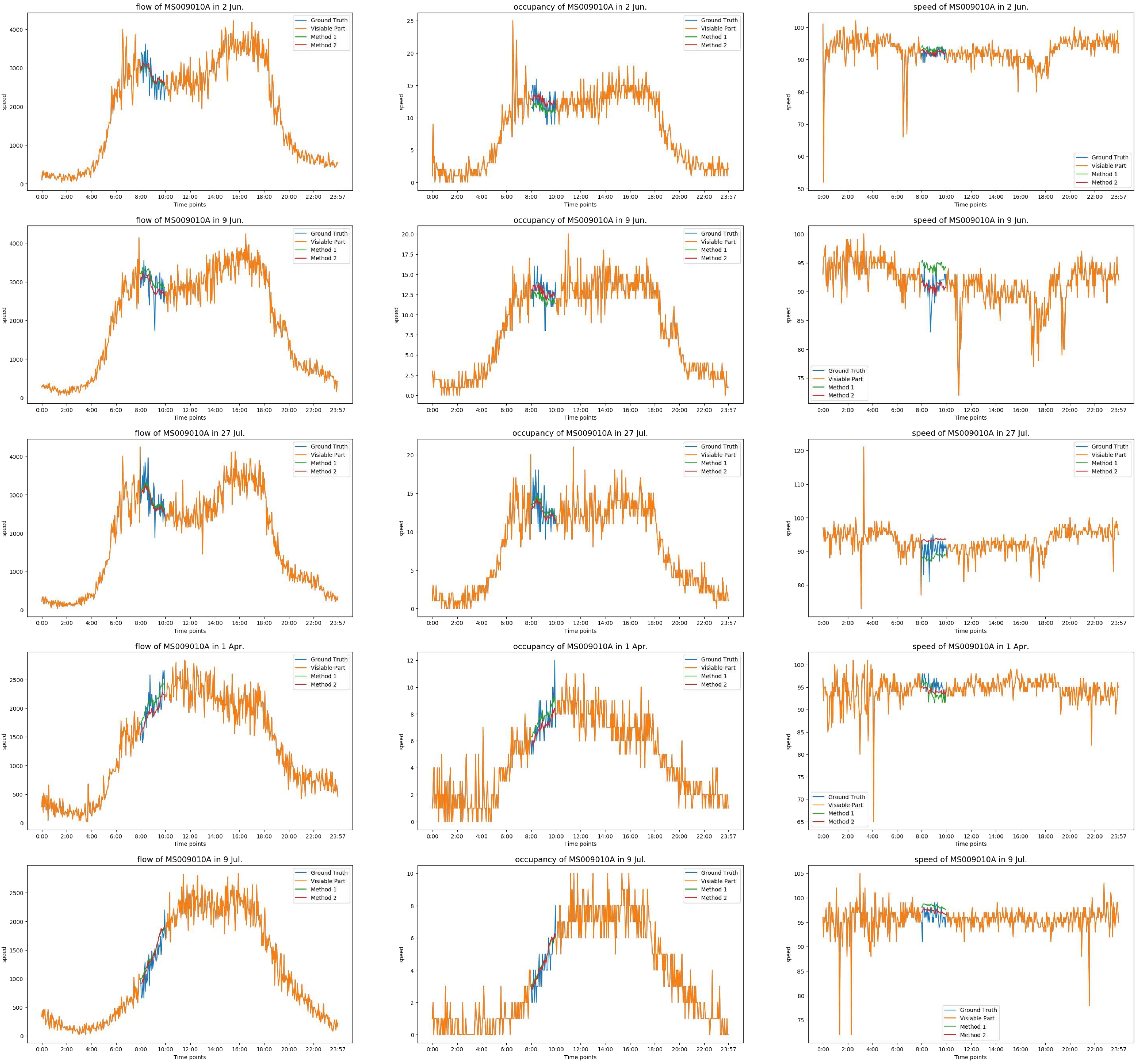}
	\caption{ \verify{Final results of data repairing among all features: flow, speed and occupancy.}}
	\vspace{-0.5cm}
	            \label{SPeed_occ_flow_repair_final_results}
\end{figure}

To better showcase the efficiency of the proposed repairing procedure, in Fig. \ref{SPeed_occ_flow_repair_final_results} we further provide various examples of repaired results across several stations along the motorway for all data streams (speed, occupancy and flow) from which we have hidden away specific data parts to replicate anomalies (we did this so we can have a ground truth - marked in blue- to compare the efficiency of the repairing procedure). The repairing procedure which we denote as Method 2 (marked in red) for this example has been also compared to a simple interpolation algorithm of only using the DPPs which is denoted as Method 1 (marked in green). The results showcase very good repairing results across all selected stations (due to lack of space this is just a random selection for visualisation purposes) under "replicated anomaly conditions" (so we can keep a ground truth for validation purposes of the approach), especially for flow and occupancy data streams; the speed repairing efficiency when using Method 1 seems to be either "over or under-estimating" speed values, while the proposed Method 2 provides better outcomes in terms of repaired data stream. There are as well exceptions, such in the case of station 10A which is over estimated but for majority of cases the proposed anomaly detection and repairing algorithm is efficient. The overall benefits of Method 2 can be observed as well from Table \ref{Overall_method_comparison} which presents the averages RMSE and standard deviation results obtained across all stations and all data points from incoming streams (flow, speed and occupancy) when using both methods.

\begin{table}[pos=tbp,width=10cm,align=\centering]
	\centering
	\begin{tabular}{ccccc}
		\toprule
		Avg RMSE & Flow & Occupancy & Speed\\
		\midrule
		Method 1 & 320.37 &1.89 &2.65\\ %
		Method 2 & 290.09 &1.62 &2.55 \\
				\midrule
			 Avg Stdev &  &  & \\
        \midrule
		 	
	  Method 1 &73.01 &0.58 &0.64 \\
	  Method 2 &54.48 &0.44 &0.83 \\

		\bottomrule
	\end{tabular}
	\caption{Average RMSE and standard deviation results obtained across all stations when using both anomaly detection methods.}
	\label{Overall_method_comparison}
\end{table}

The rest of results presented in this paper will therefore focus on presenting the deep learning results by using both repaired and non-repaired data streams to showcase impact on accuracy and model performance.  

}

\section{Experimental setup}\label{S4_Exp}

In this section we describe the set-up, the implementation of the DL models and the comparison with other state-of-art prediction models. 

\subsection{\textbf{Prediction set-up:}\\}
\label{D4_experiment_setup}

\verify{
After outlier identification and anomaly processing, we run the DL models under two main set-ups: 
\begin{enumerate}[label=\alph*)]
	\item using a raw selection of the data set without anomalies which comprises 8 months of the entire dataset (see Fig. \ref{si-subfig:missing-data-counts}). More specifically, for this set-up we use the traffic flow from $01/02/17$ to $30/04/17$ and $01/06/17$ to $31/08/17$ respectively, for training the models (6 months in total). The flow from $01/09/17$-$30/09/17$ is used for validation while the flow from $01/10/17$-$31/10/17$ is used as a test set. This was kept for consistency across all DL models. 
	\item using the entire repaired data set with the above outlier and anomaly repairing approaches. \verify{not sure how the split was finally done here}  
	
\end{enumerate}
}

\subsection{\textbf{Comparison with other models:}}\label{D4_emodel_comparison}

\textbf{Daily Profile Prediction (DPP)}: is a base model in which we use the Daily Profile (described in \cref{subsec:Sec3_2_Daily_Profile}) computed for each station and each day of the week as a predictor; 
therefore for each station, we have 7 flow curves and each curve is consisting of 480 flow values (the time interval between counts is 3min). This model therefore predicts the average traffic flow per station. 

\textbf{BPNN for separate station prediction (Sep-BPNN)}: besides the BPNN model applied over all stations, we have also applied the prediction to individual stations as well, for showcasing how the prediction would be impacted if no information on neighbouring stations would be available to the models. Therefore this model is a combination of $N$-BPNN models consuming independently historical flow of each station, each using a hidden layer of 10 neurons.  
 
\textbf{ARIMA}: the ARIMA model predicts the next value in a series as a linear combination of the past observations.
We use a default ARIMA implementation, with the hyper-parameters $ARIMA(p=2,d=1,q=0)$.
Here,
$p$ is the parameter of the autoregression (note that $p$ in ARIMA has a different meaning than $P$ in \cref{tab:notations}), $d$ is for the degree of differencing (the number of times the data have had past values subtracted) and $q$ controls the moving average. 
We have determined the values of the hyper-parameters on the validation set, through exhaustive line search in the domains $p \in \{1, .. , 5\}$, $d \in \{1, .., 5\}$, and $q \in \{0, .., 3\}$.
We used 100 flow counts as the maximum historical time horizon for training. 
Note that ARIMA can only predict one value in the future.
To predict longer time horizons we add the prediction to the time series and we roll forward to the next \ti.
\subsection{\textbf{DL implementation and hyper-parameter selection:}\\}

\textbf{Avoiding overfitting.}
Even complex learners like our DL methods can overfit training data if trained for too long.
We control overfitting using the validation data-set.
At each DL epoch (i.e. learning iteration), we learn on the training set and we measure the performance on the validation set.
We record both the performance and the trained model after each epoch.
We terminate the training when the loss function on the validation dataset has not decreased for three consecutive epochs.
We select as the \emph{best trained model} (the stored model at the epoch that achieved the lowest validation error).
In practice, the training process ends in about 20-30 epochs.

\textbf{Deep learning parameters.}
We tune the values of the DL hyper-parameters on the validation set.
We vary the batch size in the range $[20,30,40,...75, 100]$ and we obtain a value of $50$. 
Our learning rate is $0.0003$ and the weight of the $L2$ regularisation term is $10^{-8}$.
All our DL models are implemented using PyTorch~\cite{paszke2017automatic}, using the Adam optimiser which provided a better performance than SGD or AdaGrad.
 
\subsection{\textbf{Past and future prediction horizon selection}}
At a given time point $t$, the input of each DL model is the traffic flow during the past $R$ time points, and the output is the prediction of the flow at the $P^{th}$ time point in the future.
Therefore, for any given station $j$ the input is $[x^{t-R+1}_j, x^{t-R+2}_j, \ldots, x^{t}_j]$, the output is $\overline{x^{t+P}_j}$ and the training performance is measured by how close the prediction is to the recorded flow $x^{t+P}_j$.
By varying $t$ on a dataset with $n$ time points, we obtain $n-R-P+1$ pairs of inputs and outputs.
Take for example $R=2$ and $P=1$ on a dataset with 5 time points.
The above procedure generates 3 train-test sets: [$x^1_j$,$x^2_j$]:$x^3_j$, [$x^2_j$,$x^3_j$]:$x^4_j$, [$x^3_j$,$x^4_j$]:$x^5_j$, where the column separates the training vector and the desired output. 
When $R=3$ and $P=2$, we have only one combination [$x^1_j$,$x^2_j$,$x^3_j$]:$x^5_j$ -- we predict the fifth time point based on the traffic flows during the first three data points.
Our dataset contains (42,721-R-P) + (44,161-R-P) combinations (as we have two separate contiguous training periods), the validation set contains (14,401-R-P) pairs and the test set contains (14,881-R-P) pairs.
 
$R$ is an important hyper-parameter of our model -- the length of the learning past horizon -- which is tuned on the validation set in the range $R \in\{1,..,30\}$.
30 time points in the past corresponds to a 90 min past time horizon, which is more than the expected travel time along the whole M7 motorway in one direction.
Given a value of the prediction time horizon $P$, we train the model 5 times and we calculate the average accuracy on the validation dataset. 
\revA{We select as the \emph{best $R$}} the value that achieves the highest average accuracy for \revA{the} current $P$.
In \cref{R_and_P}, we focus on the relationship between $R$ and $P$ in order to answer several open questions: 
a) how much should we learn from the past to achieve best prediction results? 
b) how long in the future should we predict? 
c) is the size of the past horizon affecting the prediction results? 
d) what is the relation between $R$, $P$ \revA{and} the performances of the advanced DL models?

\subsection{\textbf{Model performance \revA{and training time}}}
We evaluate prediction performances using three widely used measures;
a) the Root Mean Square Error (RMSE), b) Mean Absolute Error (MAE) and c) Symmetric Mean Absolute Percentage Error (SMAPE). 
\revA{All models are trained on an Intel Xeon processor with 24 cores, and they take between 10 and 15 epochs to converge.
For $P=5$ and $R=5$, the training time is $219.45\pm61.6$ (sec) for CNN, $302.10\pm100.9$ for LSTM and $382.53\pm107.7$ for CNN-LSTM; more information is provided in Table \ref{tab:execution-time}.
}

\begin{figure}[pos=htbp,width=\textwidth,align=\centering]

	\centering
	\subfloat[]{%
		\includegraphics[width=0.5\textwidth]{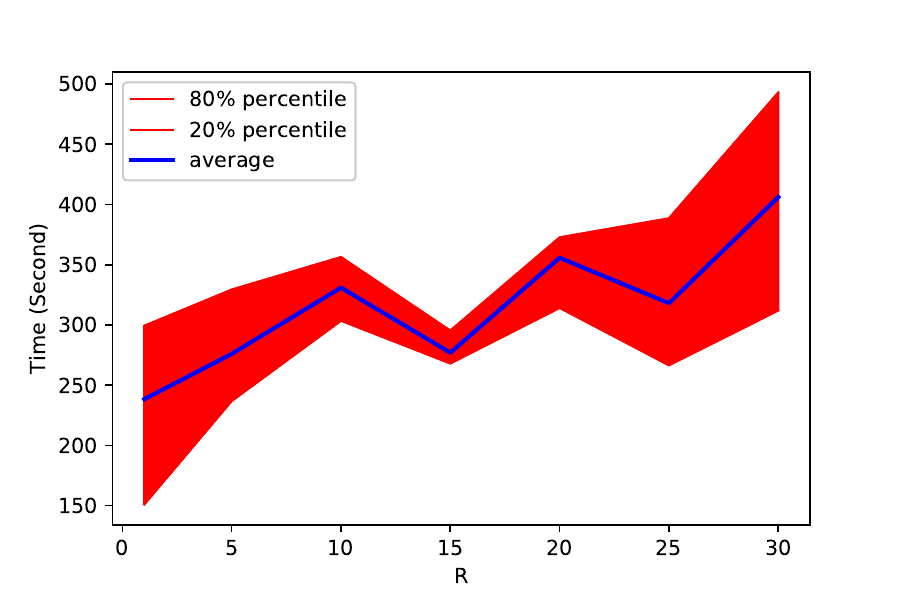}%
		\label{si-subfig:lstm-time}%
	} 
	\subfloat[]{%
		\includegraphics[width=0.5\textwidth]{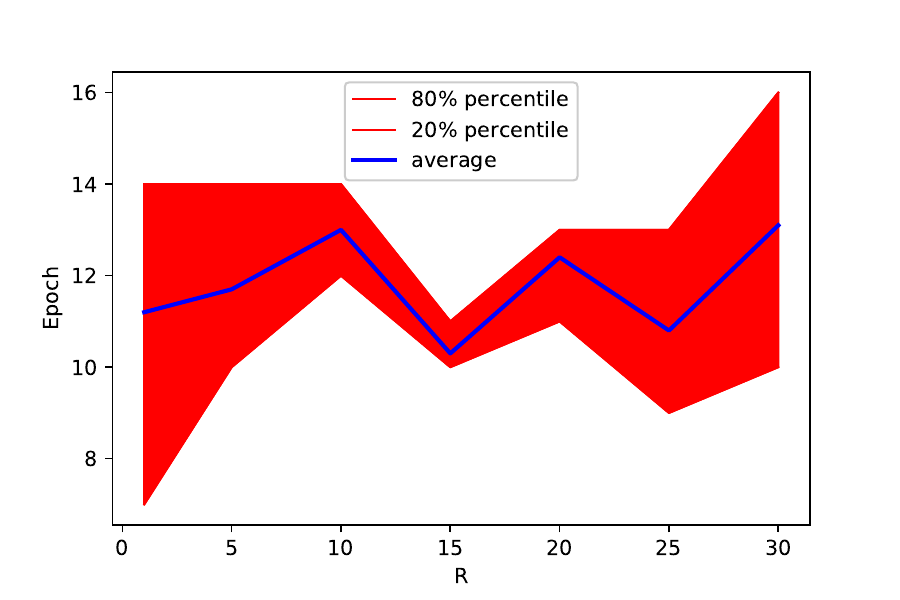}%
		\label{si-subfig:lstm-epoch}%
	}%
	\caption{
		Training time \textbf{(a)} and epochs to convergence \textbf{(b)} required by LSTM, with multiple values of $R$.
		The shaded area indicates the $20\%-80\%$ percentiles interval.
	}
	\vspace{-0.5cm}
\end{figure}

\subsection{\textbf{Training time analysis}\\}
\revA{In this subsection, we evaluate the time needed to train our advanced DL models.
The main factors that influence the training time are $R$ -- controlling the extent of the past horizon taken into account when predicting the future -- and the number of epochs required for the DL methods to converge.
$R$ is a hyperparameter of the model, larger values of $R$ require more training time per epoch.
The number of epochs until convergence are data dependent.
\cref{si-subfig:lstm-time,si-subfig:lstm-epoch} illustrate for the LSTM model the relation between the value of $R$ and the training time (\cref{si-subfig:lstm-time}), and the number of epochs until convergence and the required training time (\cref{si-subfig:lstm-epoch}). 
For each value of $R$, we repeat the training of the model 10 times, and we show the mean training time and the standard deviation.
Visibly, the training time typically increases with the value of R increase, except for $R = 15$.
We observe the mean number of epoch required to converge fluctuates between 11 and 13, which implies that the value of $R$ does not affect the number of required epochs until convergence. 
Interestingly enough, both the mean and the standard deviation of the number of epochs until convergence drops significantly for $R = 15$, which also explains the drop in training time.
The reason of this drop would require further investigation and it is part of our future work plan.
\cref{tab:execution-time} shows the mean and standard deviation for the training time for all four DL models, for $P=5$ and $R=5$.
Each training for each model is repeated 20 times.
We find that BPNN is the fastest to train, followed by CNN and LSTM.
The complex model CNN-LSTM is the slowest to train.
}

\begin{table}[pos=tbp,width=10cm,align=\centering]
	\centering
	\begin{tabular}{cccccc}
		\toprule
		& BPNN& CNN& LSTM& CNN-LSTM\\
		\midrule
		Mean & 101.190 &219.452 &302.105 &382.538\\ %
		Std &28.304 &61.610 &100.960 &107.722\\
		\bottomrule
	\end{tabular}
	\caption{The time spent on training our models [sec].}
	\label{tab:execution-time}
\end{table}
\section{Main findings}\label{S5_Results}

In this section, we present several results in terms of: the predictive comparative performance analysis for all models, the residual analysis of best DL model, the interplay between the past ($R$) and future ($P$) time horizons and best R for each P, under both setups previously described: using raw data (\cref{Prediction_performance_raw} - \cref{Best_R_for_P}) versus using anomaly-free data (\cref{Predict_performance_no_anomalies} - \cref{BEst_R_for_P_no_anomalies}). 

\subsection{DL models performance: raw data set-up}
\label{subsec:performance}

\begin{figure}[pos=htb,width=0.9\textwidth,align=\centering]
	\centering
	\includegraphics[width=12cm]{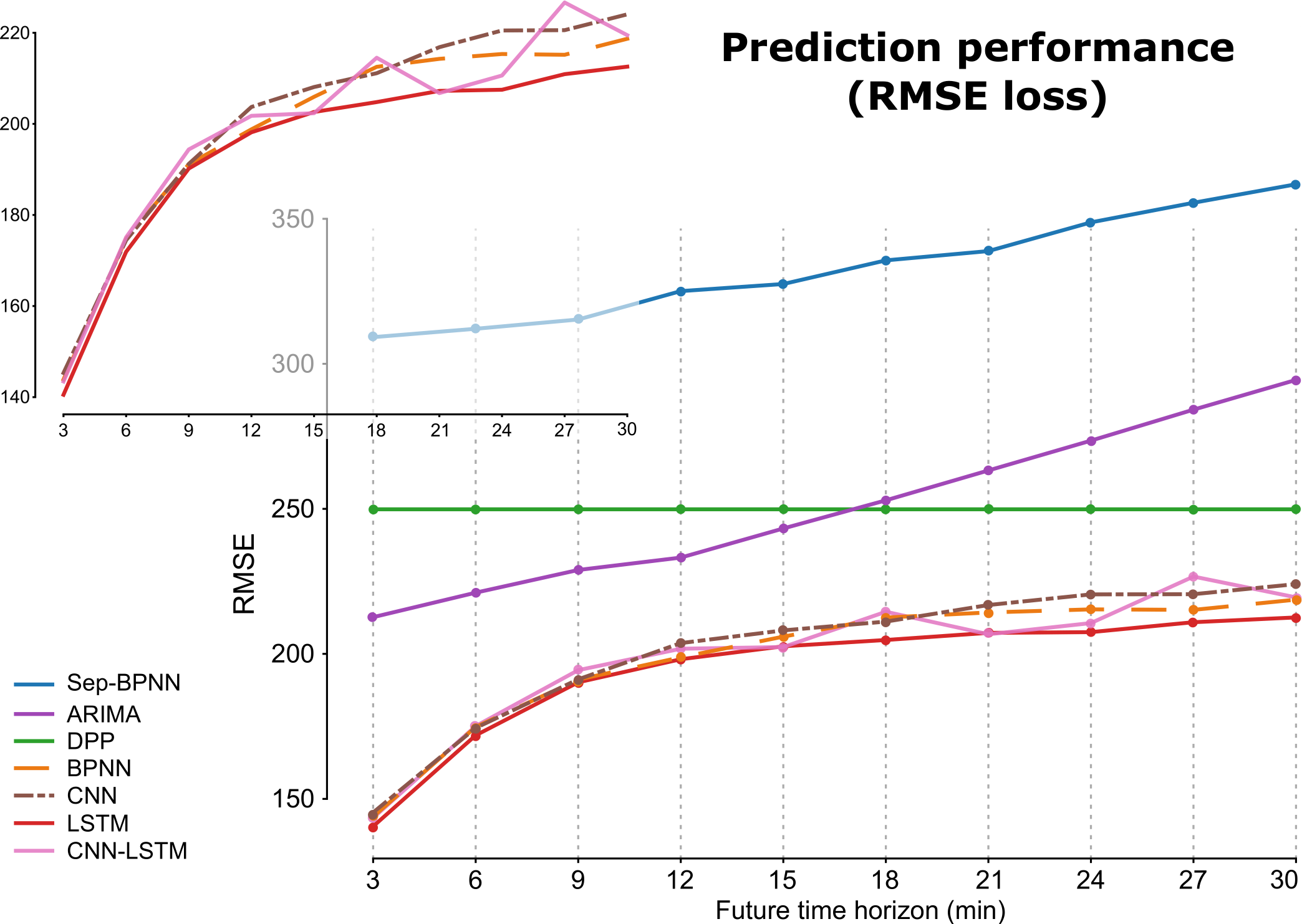}
	\caption{
		Prediction performance for all models (Oy axis), for increasing future time-horizons P (Ox axis). The zoom expands the performance of DL models. The y-axes show the RMSE of each model (lower is better).
	}
	\label{Fig_7_RMSE_all_models}
	\vspace{-0.5cm}
\end{figure}

\subsubsection{Prediction performance:}\label{Prediction_performance_raw}
\begin{figure}[pos=tb,width=\textwidth,align=\centering]

	\centering
	\subfloat[]{%
		\includegraphics[width=0.5\textwidth]{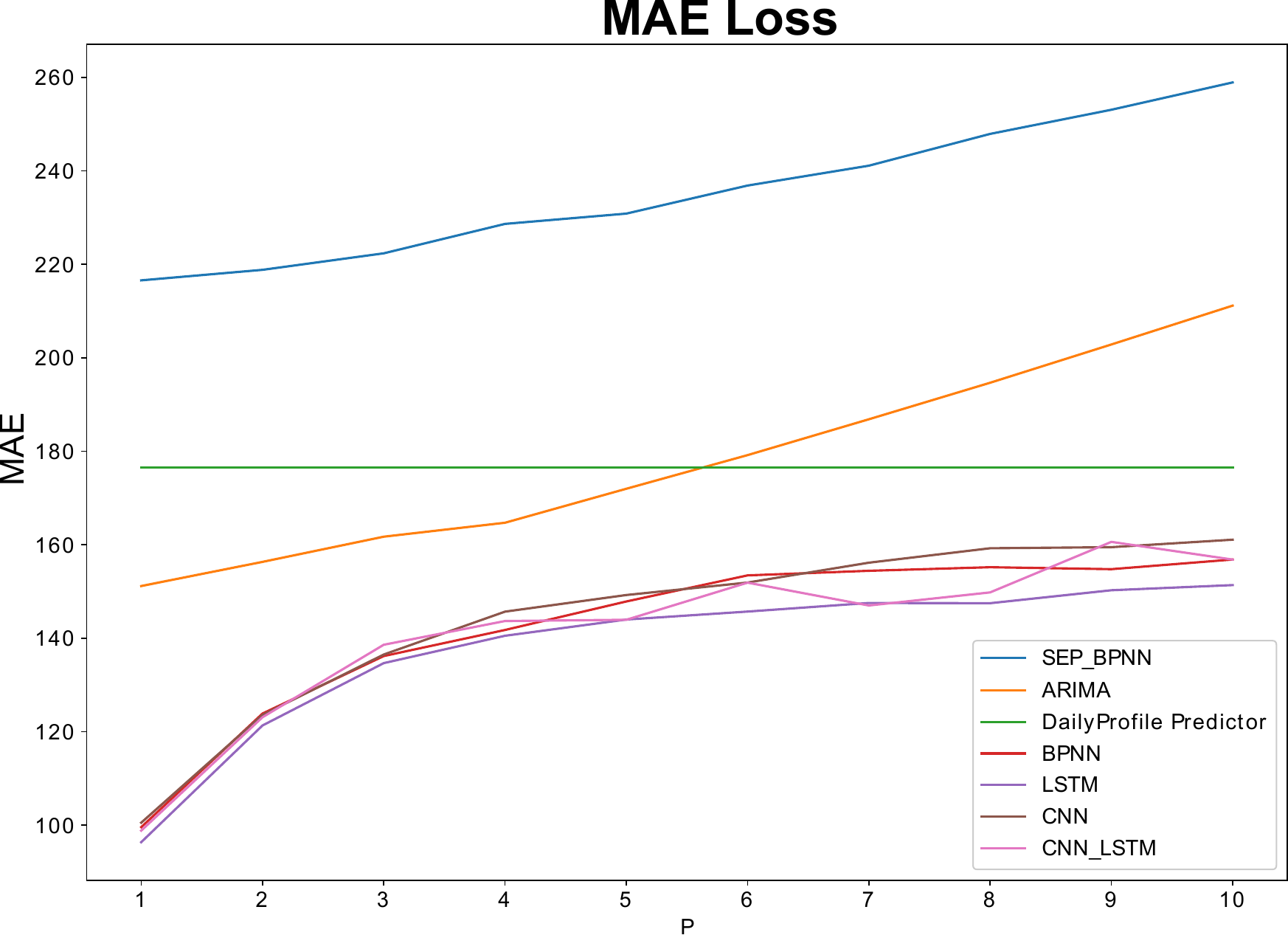}%
		\label{si-subfig:mae-loss}%
	} 
	\subfloat[]{%
		\includegraphics[width=0.5\textwidth]{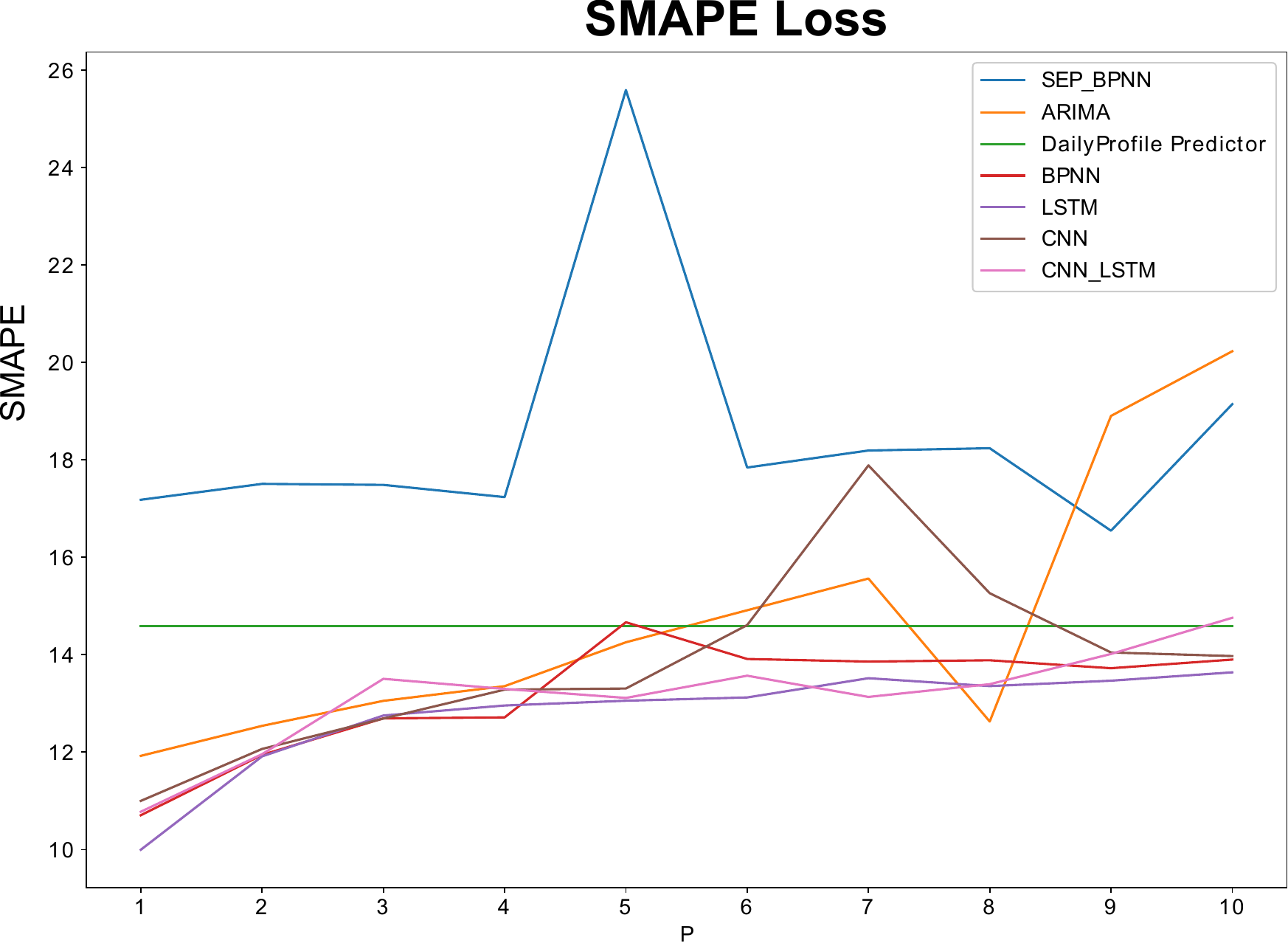}%
		\label{si-subfig:smape-loss}%
	} 
	\newline
	\caption{a) MAE loss results calculated across all models b) SMAPE error calculated for all comparative models present in this study and c) CNN-LSTM evaluation for multiple past and future horizons.}
\end{figure}

We train all DL models and baselines on the testing and validation set, with varying prediction horizons $P \in \{1,..,10\}$, where $P=1$ represents a 3min prediction, while $P=10$ is equivalent of a 30min prediction in the future.  
\cref{Fig_7_RMSE_all_models} shows the RMSE prediction error for all models.
As expected, the prediction performance of all models (except DPP) decreases as we predict further into the future. DPP (daily profile predictor) has a constant RMSE as it outputs historical averages for any data point.
Visibly, the worst performing model is the Sep-BPNN mostly because it does not incorporate the spatial and temporal correlation between the counting stations. 
The parametric model ARIMA appears to under-perform DPP for large prediction horizons, probably due to the accumulation of errors in its rolling prediction. 
The best performing models are the advanced DL models; 
LSTM outperforms all models for every $P$, followed closely by the hybrid model CNN-LSTM (which only for p=7 (21min) outperforms LSTM). 
The performances of the hybrid model fluctuate, and it is outperformed by regular CNN for $\{p\leqslant3$, $p=6$ and $p=9\}$ indicating that for our problem, a more sophisticated model does not necessarily improve performances. 
All DL models achieve similar performances for a prediction horizon lower than 12 min ($P \leq 4$), and performances stabilise after 21 min.
We also provide the additional graphics for other two performance metrics evaluated against all models. \cref{si-subfig:mae-loss,si-subfig:smape-loss} show the prediction error for all models, when measured using the MAE and SMAPE respectively. The same conclusions emerge as from the RMSE analysis presented previously. 

\subsubsection{Residuals analysis:} 
\begin{figure*}[pos=h!,width=\textwidth,align=\centering]

	\centering
	\includegraphics[width=0.99\textwidth]{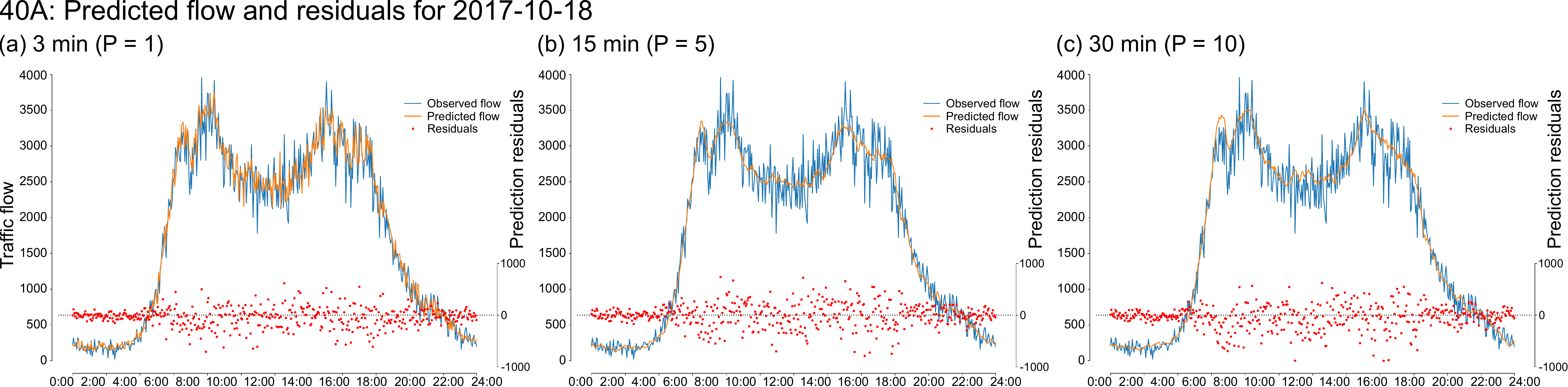}
	\caption{
		Observed and predicted traffic flow, and residuals for 3 min \textbf{(a)}, 15 min \textbf{(b)} and 30 min \textbf{(c)} for station \texttt{40A} on a weekday.}
	\label{Fig_6_LSTM_results}
	\vspace{-0.5cm}
\end{figure*} 

\cref{Fig_6_LSTM_results} shows the observed and the predicted flow by LSTM (our best performer) during a day, for station \texttt{40A}.
We also show the prediction residuals (the difference between our prediction and the real flow count information). 
The prediction results are rolled out for the next 3min (\cref{Fig_6_LSTM_results}a), 15min (\cref{Fig_6_LSTM_results}b), and 30min into the future (\cref{Fig_6_LSTM_results}c); they show good performance for short-term predictions (less than 15min) with very low residuals outside peak hours and reaching a maximum error of $10.8\%$ during AM/PM peak intervals. 
For long-term prediction, the LSTM model maintains a good prediction performance for the overall traffic flow trend, but has lower accuracies for predicting smaller traffic flow deviations from main flow profile. 
The performance of all models has been investigated as well for flow prediction during stochastic events with major disruptions on the traffic network and it will be further presented in an extended version of this work. 

\subsubsection{Interplay between past and future horizons}\label{R_and_P}

\begin{figure*}[pos=h!,width=\textwidth,align=\centering]
	\centering
	\includegraphics[width=0.99\textwidth]{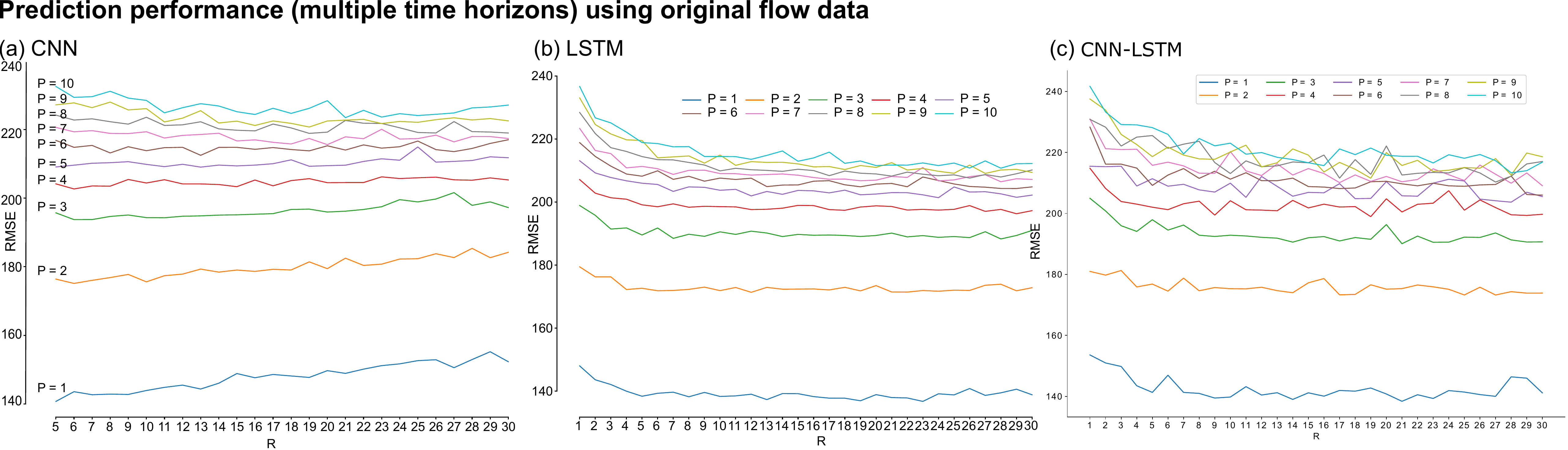}
	\caption{
		Prediction performances for multiple future time horizons for CNN \textbf{(a)}, LSTM \textbf{(b)} and CNN-LSTM \textbf{(c)} (lower is better).
	}
	\label{Fig13_14_15}
\end{figure*}

After evaluating performances, we next investigate the optimal past time horizon required by each DL model to make prediction at a given future point.
The past horizon extends up to 90min in the past ($R \leq 30$), while the explored future horizon reaches 30min in the future ($P \leq 10$).
\cref{Fig13_14_15}a and \cref{Fig13_14_15}b showcase the prediction performance over multiple time horizons for CNN and LSTM, the two most powerful DL models due to their capability to efficiently incorporate spatial and temporal features. 
Both models obtain their best RMSE for short time predictions ($P=1$), however CNN suffers a significant decrease in performance if we consider longer past horizons ($R>13=39min$).
This can be explained by the fact that CNN is designed for leveraging spatial correlations, and make less usage of temporal information. 
LTSM's performance improve with the availability of longer historical, admittedly slower for larger values of $R$.
Both models present a decreasing performance when we predict too far into the future ($P>5=15min$), and RMSE appears not to decrease significantly for large values of $P$.
\verify{CNN-LSTM behaves very similarly to LSTM (shown in \cref{Fig13_14_15}c), indicating the the hybrid model is not over-performing as various literature studies are expecting. On the other it also seems not only to under-perform the LSTM, but also to be more unstable with large various of $R$ and $P$ horizons. This was a surprising finding revealing that more complex deep learning prediction models for our case study did not bring any significant performance improvement and are also the slowest in training time. }

\subsubsection{Best $R$ for $P$:}\label{Best_R_for_P}

\begin{figure}[pos=htb,width=9cm,align=\centering]
	\centering
	\includegraphics[width=0.5\textwidth]{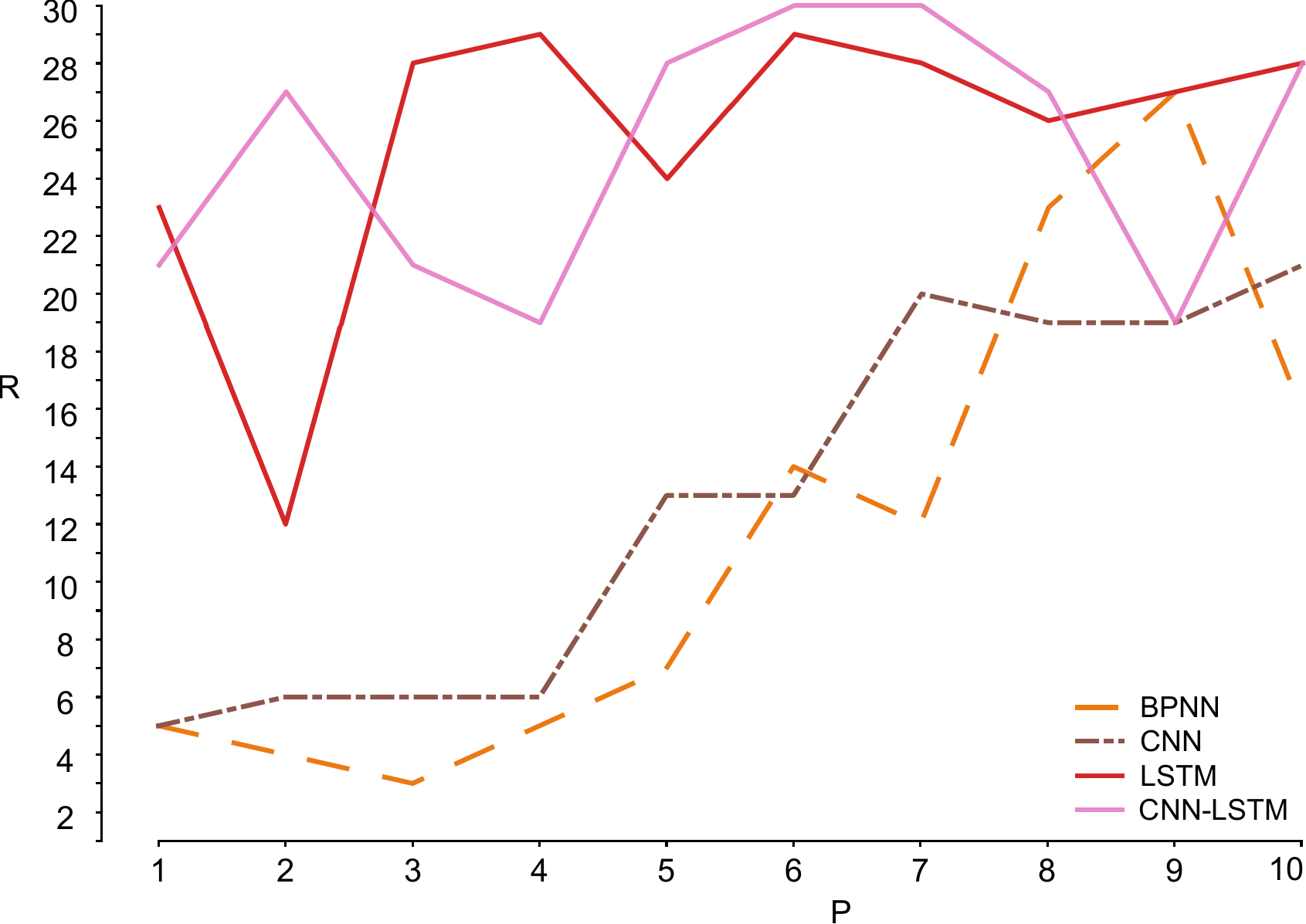}
	\caption{Best length of past time horizon ($R$) for given future time horizons ($P$).
	}
	\label{Fig_14_Best_R_and_P_pairs}
\end{figure}

To complete our analysis of the first set-up, in \cref{Fig_14_Best_R_and_P_pairs} we showcase the best past horizon dimension which is selected for each future prediction horizon, across each DL model.
We observe that LSTM and the hybrid CNN-LSTM make use of larger past time horizons even when making short-term predictions.
When predicting 9min ahead ($P=3$), the best LSTM performance leverages 69min in the past ($R=23$), whereas CNN only uses 18min in the past ($R=6$).
This results again reinforces the fact that LSTM and CNN-LSTM can learn long-term trends to make more accurate predictions.
Though not our case, it may prove problematic when long historical data is not available, in which case CNN and BPNN might provide better results.

\verify{
\subsection{DL models performance: anomaly repair set-up}\label{repaired_anomalies}

\subsubsection{Prediction performance under no anomalies:}\label{Predict_performance_no_anomalies}

The second experimental set-up was following the same design of experiments, with the only difference being the data set in use which this time consisted in the anomaly free and outlier removal process being applied. The main purpose of this set-up is to analyse and evaluate how the accuracy of such models would evolve/improve in combination with an outlier and anomaly detection procedure,as well as to understand what combination of available features would provide the best accuracy for these models. 

\begin{figure*}[pos=h!,width=\textwidth,align=\centering]
	\centering
	\subfloat[]{%
		\includegraphics[width=0.5\textwidth]{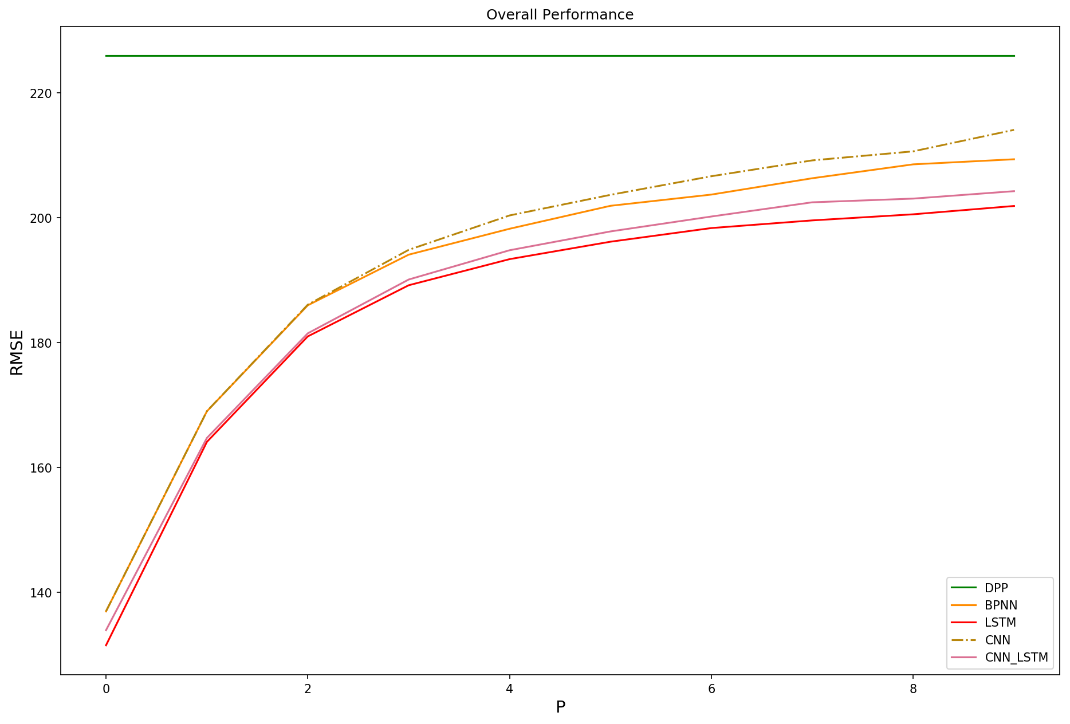}%
		\label{RMSE_DL_perform_after_cleaning}%
	}%
	\subfloat[]{%
		\includegraphics[width=0.5\textwidth]{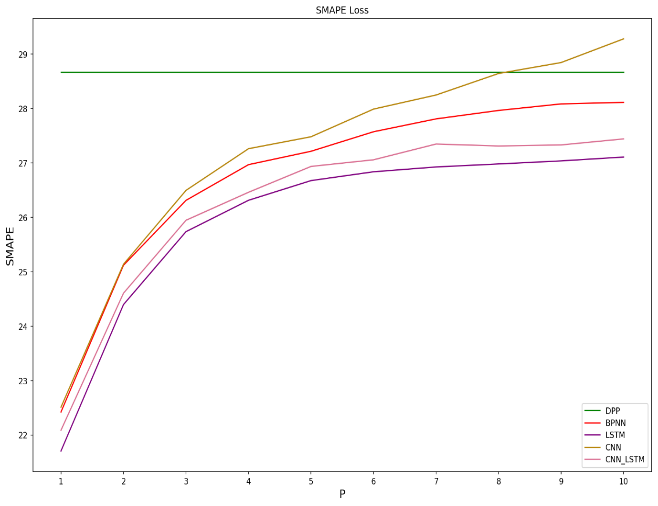}%
		\label{SMAPE_DL_perform_after_cleaning}%
	}%
	\caption{ No anomalies set-up: RMSE and SMAPE values. \verify{@Haowen please redo this figure using same colours in all figures for consistency.}}
\end{figure*}

We first start by comparing the performance of the most pertinent models (DPP, BPNN, LSTM, CNN and CNN-LSTM) in terms of RMSE and SMAPE, as represente din Fig. \ref{RMSE_DL_perform_after_cleaning} and \ref{SMAPE_DL_perform_after_cleaning}. When comparing the RMSE values, there is a significant reduction across all models (except DPP which remains constant), but more specifically in the smoothness behaviour of LSTM and CNN-LSTM with the increase of the prediction horizon, which seem to clearly over-perform the other models. The anomaly detection improvement on the performance of DL models is also backed-up by the SMAPE loss which confirms the previous findings once again.

\subsubsection{Residual analysis under no anomalies:}\label{Residuals_no_anomalies}

As LSTM maintains its over-perfoming capabilities for the second set-up as well, we also provide in Fig. \ref{LSTM_no_anomalies_with_incident_data} the comparison of prediction results for various prediction horizons ahead (3min, 15min, and 30min), this time on a main station ($22A$) which has been affected by a reported incident taking place between 5-6 PM in the afternoon (validated against incident records received as well for the same time period). As this is not an outlier situation, nor a missing record or abnormal drop in the traffic stream, the anomaly detection was not triggered to repair the data streams; however, due to previous repairs in the data streams and a better representation of the daily profiles, LSTM managed to provide better prediction results up to 15min flow prediction which adapted to the new incoming traffic flow reduction as compared to previous situation depicted in Fig. \ref{Fig_6_LSTM_results}. For longer horizon prediction LSTM manages to predict the overall trend but it highly depends on latest incoming data streams for adapting to the new changing traffic situation reflected by a potential incident.   

\begin{figure}[pos=htb,width=\textwidth,align=\centering]
	\centering
	\includegraphics[width=\textwidth]{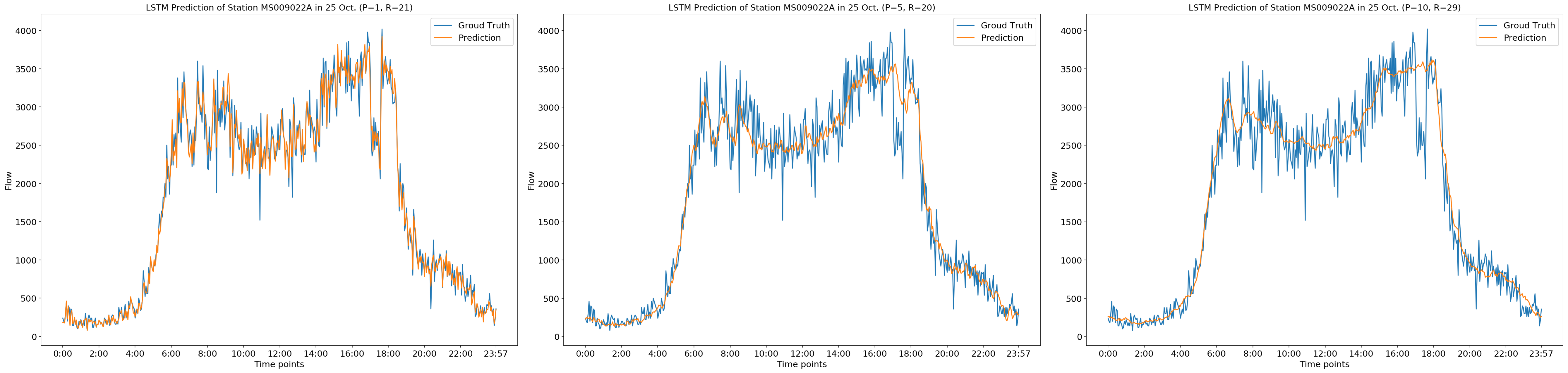}
	\caption{No anomalies set-up: LSTM observed and predicted traffic flow, and residuals for 3 min \textbf{(a)}, 15 min \textbf{(b)} and 30 min \textbf{(c)} for station \texttt{22A} on an incident day - 25th of Oct 2017.
	}
	\label{LSTM_no_anomalies_with_incident_data}
\end{figure}

\subsubsection{Past versus future horizons under no anomalies:}\label{Past_Future_no_anomalies}

\begin{figure*}[pos=h!,width=\textwidth,align=\centering]
	\centering
	\includegraphics[width=0.9\textwidth]{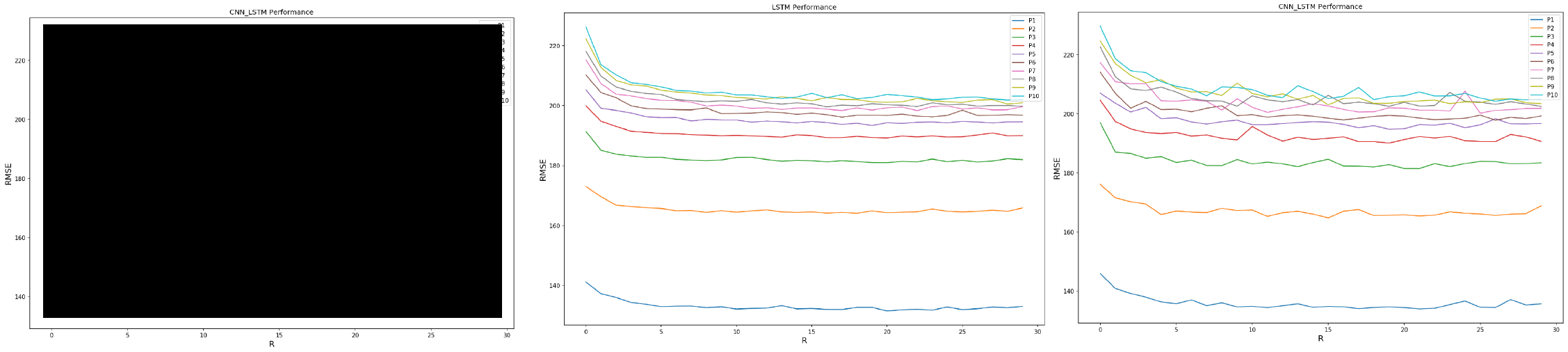}
	\caption{
		No anomalies set-up: prediction performances for multiple future time horizons for CNN \textbf{(a)}, LSTM \textbf{(b)} and CNN-LSTM \textbf{(c)} (lower is better).
	}
	\label{CNN_LTSM_hybird_no_anomalies}
\end{figure*}

Similarly to Fig. \ref{Fig13_14_15}, we also provide the performance of CNN, LSTM and hybrid CDD-LSTM under all past versus future horizon combinations, this time using the repaired and anomaly-free data set as depicted in Fig. \ref{CNN_LTSM_hybird_no_anomalies}. While the same trend in terms of accuracy evolution is maintained in this case as well with lower RMSE values than before, one important aspect to notice is the smoothness and stability of the hybrid model for longer prediction horizons. This showcases the effectiveness of the anomaly detection method on prediction accuracy.  

\subsubsection{Best R for P under no anomalies:}
\label{BEst_R_for_P_no_anomalies}

BY re-running the experiments in order to find the best past horizon R for predicting in a specific future horizon P, we have re-generated Fig. \ref{Fig_14_Best_R_and_P_pairs} as shown in Fig. \ref{Best_R_for_each_P_no_anomalies} and observed a specific convergence across all models after P=5 (15min) horizon into the future. While the trend remains almost the same for LSTM and the hybrid mode, and the BPNN and CNN, there is a significant convergent across all models which require now less historical traffic information than before: see for example that for p=8 (24min) in the future, both LSTM and BPNN require only r=16 (48min) past horizon, whereas before they would require information from r=26 and r=22 respectively. This showcases that accurate and clean data stream can help not only improving the prediction accuracy, but mostly can hep reducing the amount of historical information needed to produce similar results. 

\begin{figure}[pos=htb,width=12cm,align=\centering]
	\centering
	\includegraphics[width=0.6\textwidth]{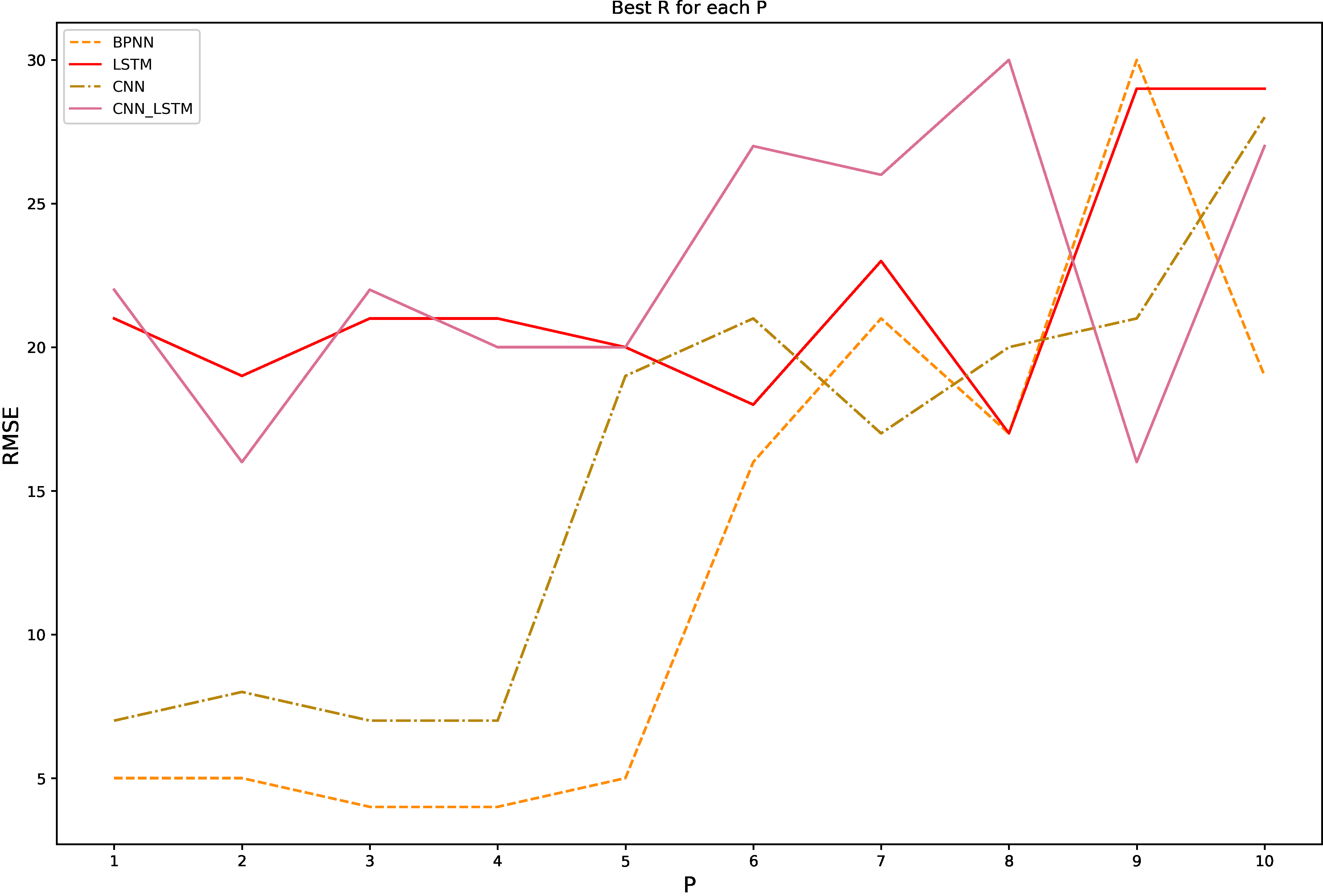}
	\caption{No anomalies set-up: Best length of past time horizon ($R$) for given future time horizons ($P$).
	}
	\label{Best_R_for_each_P_no_anomalies}
\end{figure}

\subsection{All DL training combinations on incoming feature data set}
\label{Sect_All_DL_combinations}

Last but not least, we have performed various training combinations acorss all data streams received and in Fig. \ref{All_DL_combinations} we provide all prediction results which have been obtained if we would have trained and the DL models: a) on each separate data stream (speed, flow, occupancy) or either on a combination of b) two or c) all features. The findings are interesting and one can easily observe that the best performance metrics are obtained when predicting and training mostly using the traffic flow (lowest values for MAPE = 13.25, MAE=141.036 and RMSE=193.28), followed by the case of using Flow and Occupancy (MAPE = 13.44, MAE=142.61, RMSE=194.5) and very closely the case when using all incoming data sources (MAPE = 13.68, MAE=143.61, RMSE=196.13). The worst performance seems to be obtained when using only speed to predict the traffic congestion as this is highly correlated to the flow and time of day as previously discussed from the speed-flow diagram. The results can enforce the fact that good prediction results can be obtained even if less traffic flow features can be available for DL training. An interesting future direction would be to incorporate even more features in the DL model training such as weather, incidents, public events, etc. when they will be available. 

\begin{figure}[pos=htb,width=\textwidth,align=\centering]
	\centering
	\includegraphics[width=\textwidth]{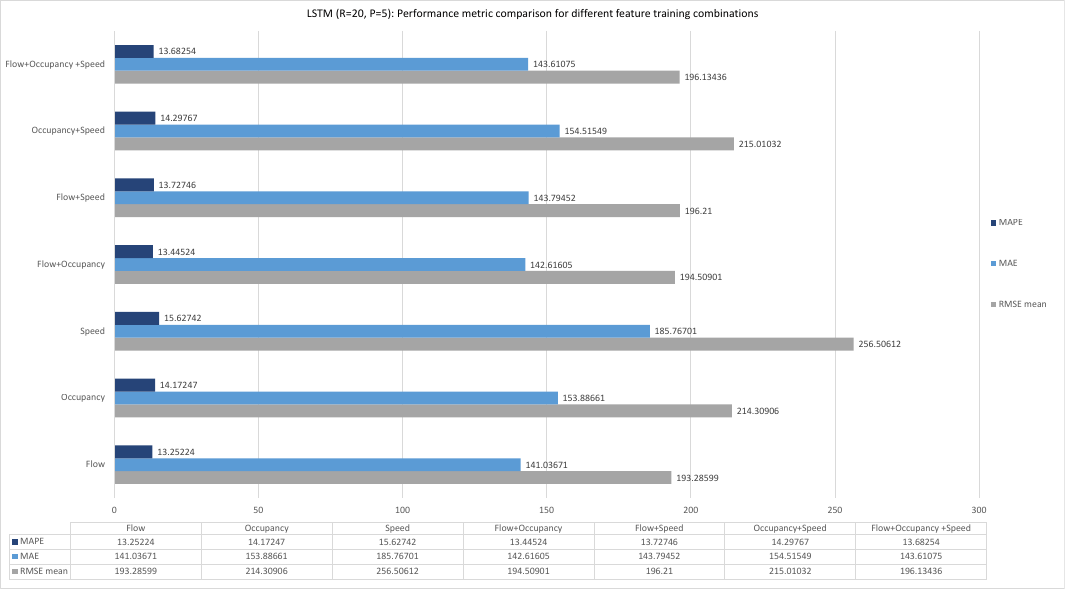}
	\caption{Performance metrics obtained by applying various DL training combinations on all incoming data streams.
	}
	\label{All_DL_combinations}
\end{figure}

}
\section{CONCLUSIONS}\label{S6_Conclusions}
This paper presents an advanced DL framework for motorway traffic flow prediction,  by chaining together data profiling and outlier identification, spatial and temporal feature generation and various DL model development.
The current approach has been applied on 36.34 million data points, to make traffic flow prediction over an entire motorway in Sydney Australia. 
The findings showcase LSTM as having the best predictive performance, despite having competed against a hybrid model combining CNN  and LSTM.
Our analysis reveals that the optimal past time horizon needs to be adapted for each DL model: LSTM and its variants learn long-term trends and require longer histories, while CNN learns spatial correlations from short histories.

\verify{Starting from the initial 3 challenges listed in \cref{S1_Intro}, we summarise the advantages of our proposed deep learning modelling: 

\begin{itemize}
	\item it provides good prediction accuracy for a large number of counting stations,
	\item its usage is based on a tailored selection of past learning horizon and future prediction horizon
	\item the hybrid CNN-LSTM model under-performed when compared to individual LSTM, which indicates that the more complex deep learning models do not improve the prediction accuracy for our motorway flow prediction study.
	\item applying an anomaly detection and outlier removal can improve significantly the prediction results, as well as the amount of historical information that is needed to train and adjust the DL models. 
\end{itemize}
}

Our \verify{future} work includes designing traffic flow-based detection methods for larger transport networks, encompassing larger areas and complex structures, an alternative that we are exploring is to explicitly incorporate the spatial relations between traffic stations using CNNs with graph structured data (see \cite{YAN2019259}, \cite{Henaff2015DeepCN}).

stochastic events which can massively disrupt the traffic flow along motorways.
For

{ \small
\section*{ACKNOWLEDGMENT}
This work has been done as part of the ARC Linkage Project LP180100114. 
The authors are highly grateful for the support of Transport for NSW, Australia. 
}

\bibliographystyle{cas-model2-names}

\bibliography{cas-refs}

\bio{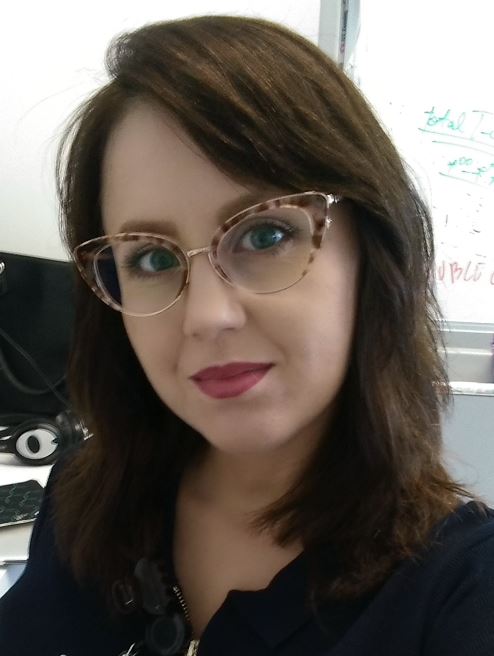}
Adriana-Simona Mihaita is a Senior Lecturer in the University of Technology in Sydney, leading the Future Mobility Lab. Her research focus is synergizing traffic simulation and optimization techniques by using machine learning to improve the traffic congestion, predict incident, while also leveraging smart analytics for connected and autonomous vehicles in a smart city environment. Dr. Mihaita holds several project management and leadership roles in various initiatives such as the Australian-Singapore Strategic Collaboration Partnership, the “Premiere’s Innovation Initiative”, the ``On-Demand Mobility'' trials in Northern Beaches in partnership with Keolis Downer, as well as the Investigation of positioning accuracy of connected vehicles operated by the Road Safety Centre in TfNSW. She is finalist in the "2019 Australian Smart Cities Award", nominated for the "2019 Prime Minister's Science Awards", and winner of the 2018 National ITS Awards.
\vspace{3mm}

\endbio

\bio{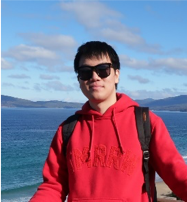}
Haowen Li is currently a research assistant at University of Technology, Sydney (UTS) and an honours-year undergraduate student from the Australian National University (ANU). He has strong programming and athematical skill. He also has experience in transportation data analysis using deep learning, face encoding learning and object detection. He has worked on several individual and group projects, including face verification about disguised as well as the impostor faces, detecting as well as recognizing house numbers from street view images.
\endbio
\vspace{3mm}

\bio{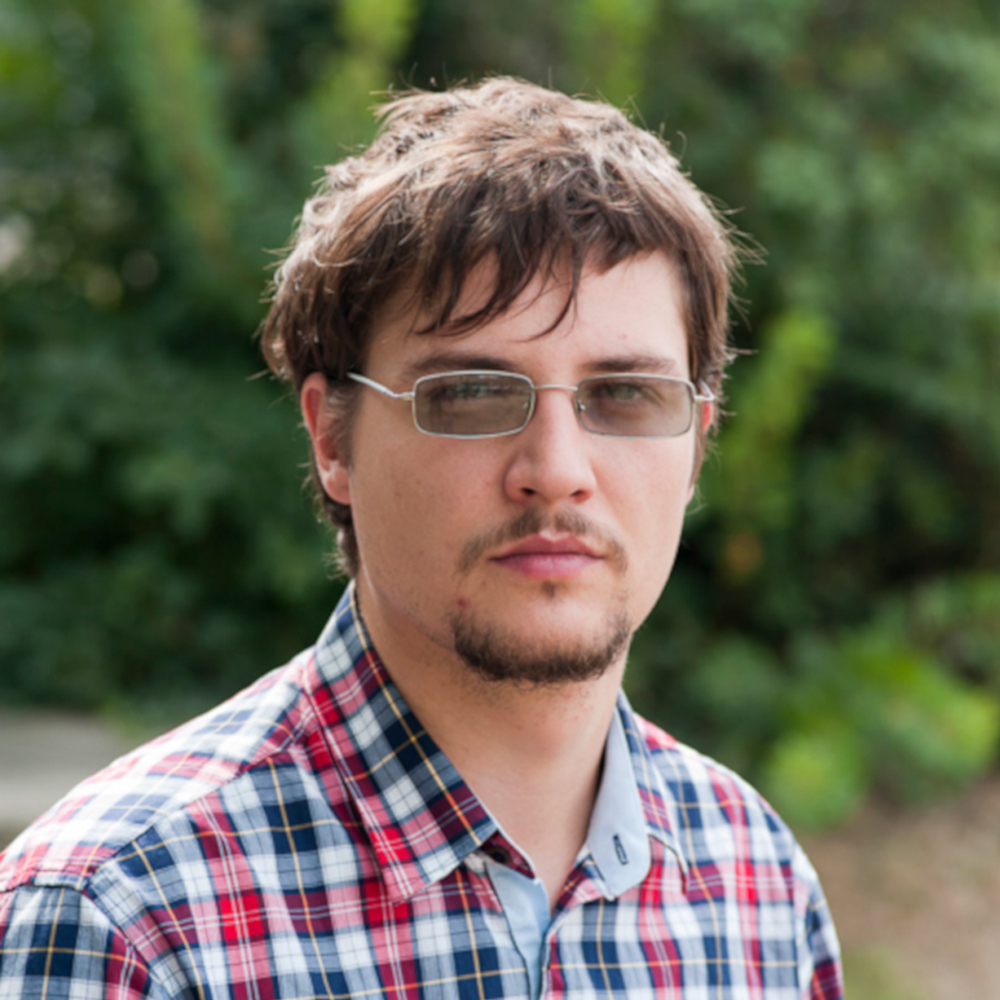}
Marian-Andrei Rizoiu is a Lecturer in Computer Science with the Faculty of Engineering and IT in the University of Technology Sydney and leader of the Behavioral Data Science group. His main research interest is to develop behavioural models for human actions online, at the intersection of applied statistics, artificial intelligence and social data science, with an interdisciplinary focus on social influence and information diffusion in online communities. He has strong data science skills and has actively collaborating with the team on the incident and congestion prediction works.
\endbio

\end{document}